\documentstyle[psfig]{cici}

\def\simg{\mathrel{%
      \rlap{\raise 0.511ex \hbox{$>$}}{\lower 0.511ex \hbox{$\sim$}}}}
\def\siml{\mathrel{%
      \rlap{\raise 0.511ex \hbox{$<$}}{\lower 0.511ex \hbox{$\sim$}}}}
\def\eq{eq$.$} \def\eqs{eqs$.$} \def\etal{et al$.$ } \def\eg{e$.$g$.$ } \def\ie{i$.$e$.$ }
\def\c2nu{\chi^2_\nu} \def\epsB{\varepsilon_B} \def\epsi{\varepsilon_i}
\def\microJy{\,{\rm \mu Jy}} \def\E0{{\it E}_0} \def\Ei{{\it E}_i}
\def\cm3{\,\rm cm^{-3}} \def\Ten53{10^{53}\, {\rm \frac{erg}{sr}}}
\def\ergsr{{\rm ergs\; sr^{-1}}}
\def\Da{\Delta \alpha} \def\deg{^{\rm o}}
\def\totc{\theta_{obs}/\theta_{core}}
\def\tc{\theta_{core}}  \def\tj{\theta_{jet}}

\def\reference{\bibitem}

\begin{document}

\title[ Jets, Structured Outflows, and Energy Injection in GRB Afterglows ]
      { Jets, Structured Outflows, and Energy Injection in GRB Afterglows: Numerical Modelling }

\author[A. Panaitescu]{A. Panaitescu \\
        2511 Speedway, Department of Astronomy, University of Texas at Austin, Austin, TX 78712 \\
        Space Science and Applications, MS D466, Los Alamos National Laboratory, Los Alamos, NM 87545 }

\maketitle

\begin{abstract}
 \begin{small}
 We investigate numerically the ability of three models (jet, structured outflow, energy injection) 
 to accommodate the optical light-curve breaks observed in 10 GRB afterglows (980519, 990123, 990510, 991216, 
 000301c, 000926, 010222, 011211, 020813, and 030226), as well as the relative intensities of the radio, optical, 
 and $X$-ray emissions of these afterglows.  
 We find that the jet and structured outflow models fare much better than energy injection model in accommodating 
 the multiwavelength data of the above 10 afterglows. For the first two models, a uniform circumburst medium 
 provides a better fit to the optical light-curve break than a wind-like medium with a $r^{-2}$ stratification.
 However, in the only two cases where the energy injection model may be at work, a wind medium is favoured
 (an energy injection is also possible in a third case, the afterglow 970508, whose optical emission exhibited
 a sharp rise but not a steepening decay). 
 The best fit parameters obtained with the jet model indicate an outflow energy of $2-6 \times 10^{50}$ ergs
 and a jet opening of $2\deg-3\deg$. Structured outflows with a quasi-uniform core have a core angular size
 of $0.7\deg-1.0\deg$ and an energy per solid angle of $0.5-3\times 10^{53}\,\ergsr$, surrounded 
 by an envelope where this  energy falls-off roughly as $\theta^{-2}$ with angle from the outflow axis, requiring 
 thus the same energy budget as jets. Circumburst densities are found to be typically in the range $0.1-1\cm3$, 
 for either model. 
 We also find that the reverse shock emission resulting from the injection of ejecta into the decelerating 
 blast wave at about 1 day after the burst can explain the slowly decaying radio light-curves observed for the 
 afterglows 990123, 991216, and 010222.
 \end{small}
\end{abstract}


\section{Introduction}

  Since the prediction of radio (Paczy\'nski \& Rhoads 1993) and optical transients (M\'esz\'aros \& Rees 1997) 
associated with GRBs, more than 100 afterglows have been observed (including $X$-ray transients).  
Good monitoring in all three frequency domains has been achieved for about 20 afterglows; 
the radio, optical, and $X$-ray flux was observed to decay as a power-law, $F_\nu \propto t^{-\alpha}$ 
($\alpha > 0$), confirming the expectations. For 10 of these well-observed afterglows -- 980519, 990123, 
990510, 991216, 000301c, 000926, 010222, 012111, 020813, 030226 -- the optical light-curve decay exhibits 
a steepening at about 1 day after the burst, with an increase $\Da$ in the temporal index as low as 0.4 
(afterglow 991216) and as high as 1.8 (afterglow 000301c). 
Optical light-curve breaks have or may have been observed in other afterglows; they are not included in
the sample of afterglows modelled in this work because those afterglows have been adequately monitored at 
only one optical frequency.

 With the exception of the afterglow 010222, for which one, late (10 days) measurement suggests that a break 
occurred also in the $X$-ray light-curve, currently available $X$-ray observations do not extend sufficiently 
before and after the optical break time to prove that the light-curve break is achromatic. Within the measurement
uncertainties, a single power-law fits well the decay of all adequately monitored $X$-ray afterglows:
990123, 990510, 991216, 000926 (and 010222 until the last measurement). 

 Breaks, in the form of peaks, have been observed in the radio emission of all 10 afterglows above, usually at 
$\sim 10$ days after the burst, \ie often occurring after the optical light-curve break. Given that the radio 
and optical breaks are not simultaneous, they cannot both arise from the dynamics of the afterglow, the structure 
of the GRB ejecta, or some property of the circumburst medium (CBM), as these mechanisms should yield achromatic 
breaks. Furthermore, given that, before their respective breaks, the radio emission rises while the optical falls-off, 
these temporal features cannot be due to the passage of the same afterglow continuum break through the observing band. 

 With the possible exception of the afterglow 990123, there is no evidence for an evolution (softening) of the 
optical continuum across the light-curve break for the above 10 afterglows (\eg Panaitescu 2005). Furthermore, 
the only spectral feature whose passage could yield the large steepening $\Da$ observed in some cases -- the peak 
of the forward shock (FS) continuum\footnotemark --
\footnotetext {This is the smallest of the synchrotron frequency $\nu_i$ corresponding to the typical post-shock 
               electron energy (which we call "injection frequency") and to the electrons which cool radiatively
              on a dynamical timescale (the cooling frequency $\nu_c$)}
should yields a rising or flat optical light-curve before the break, contrary to what is observed. 
Therefore, it is the radio peak which should be attributed to a spectral break crossing the observing domain.
That spectral break should be the injection frequency $\nu_i$, which, for reasonable shock microphysical 
parameters, should reach the radio at around 10 days. Indeed, for the afterglow 991208, there is observational 
evidence (Galama \etal 2000) that the peak of the afterglow continuum crosses the 10--100 GHz domain at that time. 
Further evidence for the passage of a spectral break through radio is provided by that the peak time of the radio 
flux of the afterglow 030329 increases with decreasing observing frequency (Frail \etal 2005). 

 Today, the generally accepted reason for the optical light-curve break is the narrow collimation of the GRB ejecta.
As predicted by Rhoads (1999), if the GRB ejecta are collimated, then the afterglow light-curve should exhibit 
a steepening when the jet begins to expand sideways. More than half of the steepening $\Da$ is due by the finite 
angular opening of the jet: as the GRB remnant is decelerated (by sweeping-up of the CBM) and the relativistic 
Doppler beaming of the afterglow emission decreases, an ever increasing fraction of the emitting surface becomes 
visible to the observer; when the jet edge is seen, that fraction cannot increase any longer and the afterglow 
emission exhibits a faster decay (Panaitescu, M\'esz\'aros \& Rees 1998). 
Because it arises from the blast wave dynamics, a light-curve break should also be present at radio wavelengths at
the same time as the optical break. However, radio observations before 1 day are very scarce and strongly affected 
by Galactic interstellar scintillation (Goodman 1997) until after 10 days, thus they cannot disentangle the 
jet-break from that arising from the passage of the FS peak frequency.

 Another mechanism for the optical light-curve breaks has been proposed by Rossi, Lazzati \& Rees (2002)
and Zhang \& M\'esz\'aros (2002): if GRB outflows are endowed with an angular structure (\ie non-uniform
distribution of the ejecta kinetic energy with direction), as first proposed by M\'esz\'aros, Rees \& Wijers (1998),
then a steepening of the afterglow decay would arise when the brighter, outflow symmetry axis becomes visible to 
the observer. In this model, the stronger the angular structure is, the larger the break magnitude $\Da$ should be.

 A third mechanism for breaks rests on the proposal of Paczy\'nski (1998) and Rees \& M\'esz\'aros (1998) 
that the FS energizing the CBM could be refreshed by the injection of a substantial energy through 
some delayed ejecta which were released at the same time with the GRB-producing ejecta, but had a smaller
Lorentz factor, or were ejected sometime later, and which catch up with the decelerating FS during the 
afterglow phase. Fox \etal (2003) have proposed that the early (0.003--0.1 day), slow decay of the optical 
emission of the afterglow 021004 is caused by such an injection process. In this scenario, when the energy 
injection episode ends, the FS deceleration becomes faster and the afterglow emission should exhibit a steepening. 

 Note the various origins of the afterglow light-curve break in each model. In the {\bf Jet} model, the break 
is caused by the changing outflow dynamics when the jet starts to spread and by the outflow's geometry. 
In the {\bf Structured Outflow (SO)} model the origin is, evidently, the outflow's anisotropic surface brightness. 
For both these models, special relativity effects play an important part. In the {\bf Energy Injection (EI)}
model, the break originates in the altered outflow dynamics at the time when the energy injection subsides.

 The purpose of this work is to compare the ability of these three models in accommodating \\
$i)$ the shape of the light-curve breaks observed in the optical emission of the afterglows
     980519, 990123, 990510, 991216, 000301c, 000926, 010222, 012111, 020813, 030226 and  \\
$ii)$ the relative intensity of the radio, optical, and $X$-ray emissions of these afterglows,  \\
for either a uniform (\ie homogeneous) CBM, or one with a $r^{-2}$ density radial stratification, 
as expected if GRBs progenitors are massive stars (Woosley 1993, Paczy\'nski 1998). 

 For the first task above, we performed an analytical test of the models (see table 2 of Panaitescu 2005), 
based on comparing  \\
$i)$ the pre- and post-break optical light-curve decay indices, $\alpha_1$ and $\alpha_2$, \\
$ii)$ those in the $X$-rays, \\
$iii)$ the slopes $\beta_o$ and $\beta_x$ of the power-law optical and $X$-ray continua 
      ($F_o \propto \nu^{-\beta_o}$ and $F_x \propto \nu^{-\beta_x}$), and  \\
$iv)$ the optical--to--$X$-ray spectral energy distribution (SED) slope, 
      $\beta_{ox} = \ln (F_x/F_o)/\ln (\nu_x/\nu_o)$,  \\
with the relations among them expected for each model. 
We note that, in the framework of the relativistic fireballs (M\'esz\'aros \& Rees 1997), the optical and 
$X$-ray decay indices and SED slopes are tightly connected, as only one continuum feature, the cooling 
frequency $\nu_c$ (Sari, Narayan \& Piran 1998), can be between these domains at the times when observations 
were usually made (0.1--100 days), and that the SED slope increases by a fixed amount, $\delta \beta = 1/2$, 
across this spectral break.

 Including the radio afterglow emission in the analytical test of the three break models is less feasible 
and often unconstraining. First, that the afterglow radio flux is modulated by diffractive and refractive 
interstellar scintillation makes it difficult to determine accurately the radio SED slope (see figs. 4 and 
5 of Frail, Waxman \& Kulkarni 2000a for the best monitored radio afterglow -- 970508). 
Second, after the injection frequency has fallen below the radio (\ie during the decay phase of the radio
light-curve), we do not expect, in general, any spectral break to be between radio and optical, hence the 
radio and optical light-curve indices should be the same. This is, indeed, the case for most afterglows;
nevertheless, there are a few troubling exceptions: over 1--2 decades in time, the radio emission of the 
afterglows 991208, 991216, 000926, and 010222 exhibits a much shallower decay than that observed at optical
wavelengths at the same time or prior to the radio decay. An analytical investigation of the various possible 
ways to decouple the radio and optical light-curves (Panaitescu \& Kumar 2004) has led to the conclusion that, 
the anomalous radio decay is due to a contribution from the reverse shock to the radio emission. 

  However, radio observations provide an indirect constraint on them because the flux and epoch of the radio peak 
determine the FS synchrotron peak flux, injection frequency, and self-absorption frequency. These three spectral 
properties constrain the afterglow parameters pertaining to the outflow dynamics (energy per solid angle, jet 
initial opening, medium density) and emission (magnetic field strength, typical post-shock electron energy), 
\ie parameters which determine the shape and epoch of the optical light-curve break, as well as the location 
of the cooling frequency. The best way to take into account the constraints arising from the radio emission and 
to test fully the three break models is to calculate numerically the afterglow dynamics and emission, and to fit 
all the available measurements. Data fitting also allows the determination of the various model parameters.

\section{Models Description}

 The basic equations employed in our numerical calculations of the outflow dynamics are presented in 
Panaitescu \& Kumar (2000) for spherical outflows, Panaitescu \& Kumar (2001) for the Jet model, Panaitescu 
\& Kumar (2003) for the SO model (structured outflows), and Panaitescu, M\'esz\'aros \& Rees (1998) 
for the EI model (energy injection). The basic equations for the calculation of the afterglow spectral
features (synchrotron and inverse Compton peak fluxes, absorption, injection, and cooling frequencies), 
and of the afterglow emission at any wavelength are given in Panaitescu \& Kumar (2000, 2001). Equations 
for the spectral characteristics can be also found in Sari \etal (1998), for the dynamics and emission from 
spherical blast-waves interacting with a uniform medium in Waxman, Kulkarni \& Frail (1998), Granot, Piran \& 
Sari (1999), Wijers \& Galama (1999) and for a wind medium in Chevalier \& Li (2000). The dynamics of
blast-waves with energy injection is also treated in Sari \& M\'esz\'aros (2000). Rhoads (1999) provided 
a detailed treatment of the dynamics and emission from jets interacting with uniform media (see also
Sari, Piran \& Halpern 1999).

\subsection{Jets}

 The jet dynamics is determined by the initial jet opening $\theta_{jet}$, the ejecta initial kinetic 
energy per solid angle $\E0$ (or, equivalently, by the jet energy $E_{jet} = \pi \theta_{jet}^2 \, \E0$),
and the CBM particle density $n$ or, in the case of a wind of a massive star, by the ratio of the mass-loss
rate to the wind speed, $(dM/dt)/v$. For convenience, the latter parameter is given normalized to 
a mass-loss rate of $10^{-5}\, M_\odot \, {\rm yr}^{-1}$ at a speed of $10^3 {\rm km \, s^{-1}}$, 
\ie for a wind
\begin{equation}
 n(r) = 0.3\, A_* r_{18}^{-2} \;\; {\rm cm^{-3}} \,
\end{equation}
where $r$ is the FS radius, and we used the usual notation $Q_n = Q({\rm cgs\, units})/10^n$.

 The deceleration of the jet, as it sweeps-up the CBM, is calculated assuming that the post-shock gas
has the same internal energy per mass (\ie temperature) as that immediately behind the FS, which is
equal to the bulk Lorentz factor of the shocked CBM, $\Gamma$. The lateral size of the jet is assumed
to increase at the co-moving sound speed, and the kinetic energy per solid angle is approximated as
uniform during the spreading. Radiative (synchrotron and inverse Compton) losses are calculated from 
the electron distribution and magnetic field strength.

 In the Jet model, the light-curve break is seen when the jet edge becomes visible to the observer.
For an observer located close to the jet axis, this time can be approximated by $\Gamma (t_{jet}) =
\theta_{jet}$, using the initial jet opening and ignoring the lateral spreading that occurred until
the jet-break time. If the FS dynamics were adiabatic, then energy conservation would lead to
\begin{equation}
  \Gamma (t)= 8.6\; (E_{0,53}/n_0)^{1/8} [t_d/(1+z)])^{-3/8} 
\label{Gs0}
\end{equation}
for a uniform medium and
\begin{equation}
  \Gamma (t)= 15\; (E_{0,53}/A_*)^{1/4} [t_d/(1+z)])^{-1/4} 
\end{equation}
for a wind, where $t_d$ is the observer time in days. Then the jet-break time is given by
\begin{equation}
  t_{jet} \simeq 0.7 (z+1) (E_{0,53} n_0^{-1} \theta_{jet,-1}^8)^{1/3} \; {\rm d}
\label{tjs0}
\end{equation}
for a uniform CBM and
\begin{equation}
  t_{jet} \simeq 5 (z+1) E_{0,53} A_*^{-1} \theta_{jet,-1}^4 \; {\rm d}
\label{tjs2}
\end{equation}
for a wind.
Given that observer locations off the jet axis (but within $\theta_{jet}$) have a small effect on the
resulting afterglow light-curves (Granot \etal 2002), we consider that, in the Jet model, the observer
is always on the jet axis. Furthermore, we ignore the possible existence of a counter-jet whose emission
would become visible to the observer when the semi-relativistic dynamics sets, \ie it would affect
the radio afterglow emission beyond 100 days after the burst.  

 At the jet-break time, the afterglow decay index at observing frequency $\nu$ steepens from that 
corresponding to a spherical outflow (Sari \etal 1998):
\begin{equation}
 \alpha_1 = \frac{1}{4} \cdot  \left\{ \begin{array}{ll} 
           3p-3 \;, & \nu < \nu_c \;\; \& \;\; {\rm unif\; CBM} \\ 
           3p-2 \;, & \nu_c < \nu \;\; \& \;\; {\rm any\; CBM}  \\  
           3p-1 \;, & \nu < \nu_c  \;\; \& \;\; {\rm wind\; CBM}
           \end{array} \right.  \;,
\label{a1sph}
\end{equation}
to $\alpha_2 = p$ (Rhoads 1999). In equation (\ref{a1sph}), $p > 0$ is the power-law index of the post-shock 
electron distribution with energy (equation [\ref{p}]).

\subsection{Structured Outflows}

 The dynamics of structured outflows is calculated similarly to that of jets. To track the lateral fluid 
flow and the change of the kinetic energy per solid angle, the outflow surface is divided in infinitesimal 
rings and we consider that each ring spreads at a rate proportional to the local sound speed and to the ring 
width. This prescription for lateral flow reduces to that given above for a jet if the outflow is uniform
and collimated.

 In this work we consider only axially symmetric power-law outflows, whose angular distribution of the ejecta 
kinetic energy per solid angle $E$ is given by
\begin{equation}
 E (\theta < \theta_{core}) = \E0 \;,\; 
 E (\theta > \theta_{core}) = \E0 (\theta/\theta_{core})^{-q} \;,
\label{Eq}
\end{equation}
with $q > 0$. The uniform core is used to avoid a diverging outflow energy for $q > 2$.
This power-law structure is sufficiently complex, given the constraining power of the available data.
Given that, in this model, the light-curve break is due to the brighter core becoming visible, the observer 
location $\theta_{obs}$ relative to the symmetry axis is a crucial parameter in determining the break time
(Rossi \etal 2002), which is the same as given in equations [\ref{tjs0}] and [\ref{tjs2}] but with $\theta_{obs}$ 
in place of $\theta_{jet}$. 

 Analytical results for the light-curve pre- and post-break decay indices can be obtained if the observer
is located within the core ($\theta_{obs} < \theta_{core}$), but only numerically for outer locations\footnotemark.
\footnotetext{For this reason, in our previous analytical assessment of the three models (Panaitescu 2005), 
              we have restricted our attention only to the $\theta_{obs} < \theta_{core}$ case, which
              yields breaks with a smaller magnitude $\Da$}
In the former case, the light-curve break occurs when the edge of the uniform core becomes visible to the 
observer, and the light-curve decay index at a frequency $\nu > \nu_i$ increases from that given in equation 
(\ref{a1sph}) to 
\begin{equation}
 \alpha_2 = \frac{1}{4-\frac{1}{2}q} \cdot \left\{  \hspace*{-2mm} \begin{array}{ll} 
                       3p-3+\frac{3}{2}q  &  \nu < \nu_c \\ 
                       3p-2+q             &  \nu_c < \nu           \end{array} \right. \;,
\label{a2s0}
\end{equation}
for a homogeneous medium and
\begin{equation}
 \alpha_2 = \frac{1}{4-q} \cdot \left\{  \hspace*{-2mm} \begin{array}{ll} 
                       3p-1-\frac{1}{2}q(p-1)  &  \nu < \nu_c \\ 
                       3p-2-\frac{1}{2}q(p-2)  &  \nu_c < \nu      \end{array} \right. \;,
\label{a2s2}
\end{equation}
for a wind medium (Panaitescu \& Kumar 2003), provided that the structural parameter $q < \tilde{q}$, 
where, for a uniform medium, $\tilde{q} = 8/(p+4)$ if $\nu < \nu_c$ and $\tilde{q} = 8/(p+3)$ if $\nu > \nu_c$.
For a wind, the two values of $\tilde{q}$ are swapped. For $q < \tilde{q}$, the emission from the outflow 
envelope ($\theta > \theta_{core}$) is brighter than that from the core, and sets the light-curve post-break 
decay index.  For $q > \tilde{q}$, the core dominates the afterglow emission and the post-break decay index is
\begin{equation}
 \alpha_2 = \alpha_1 + \left\{ \begin{array}{ll} 3/4 & {\rm unif\;CBM} \\ 1/2 & {\rm wind\;CBM} \end{array} \right.\;,
\end{equation}
\ie $\Da$ is just the steepening produced by the finite angular extent of the emitting outflow. In this case
the structure outflow model reduces to a that of a jet with sharp edges and no sideways expansion.

 For observers located outside the core, the break magnitude $\Da$ can be larger (see figs. 2, 3, and 4 in 
Panaitescu \& Kumar 2003).

\subsection{Energy Injection}

 The injection of energy in the FS has two effects: it mitigates the FS deceleration and generates
a reverse shock (RS) which energizes the incoming ejecta and contributes to the radio afterglow emission.

 For ease of interpretation, we consider an energy injection which is a power-law in the observer time:
\begin{equation}
 d\Ei/dt (t < t_{off}) \propto t^{e-1} \;,\; d\Ei/dt (t > t_{off}) = 0  \;, 
\end{equation}
where $d\Ei/dt$ is the rate of the influx of energy per solid angle. Evidently, this injection has an 
effect on the afterglow dynamics only if the total added energy, $\Ei$, is comparable or larger with 
that existing in the afterglow after the GRB phase, $\E0$. In this case, the light-curve decay index 
during the injection process is
\begin{equation}
 \alpha_1 =  \frac{1}{4}(3-e)p - \frac{1}{4}(1+e) \cdot \left\{ \begin{array}{ll} 
                 3 \;, & \nu < \nu_c \;\&\; {\rm unif\; CBM} \\ 
                 2 \;, & \nu_c < \nu \;\&\; {\rm any\; CBM}  \\
                 1 \;, & \nu < \nu_c \;\&\; {\rm wind\; CBM}    \end{array} \right.  \;,
\label{e}
\end{equation}
(Panaitescu 2005). After the energy injection subsides, the post-break index $\alpha_2$ has the value
given in equation (\ref{a1sph}) (for an adiabatic, spherical blast-wave).

 Numerically, the energy injection is modelled by first calculating the dissipated energy and that added as 
kinetic energy during the collision between an infinitesimal shell of delayed ejecta and the FS. The partition
of the incoming ejecta energy is determined from energy and momentum conservation, and depends only on the
ratio $\Gamma_i/\Gamma$, with $\Gamma_i$ being the Lorentz factor of the incoming ejecta. Numerically,
this factor is obtained from the kinematics of the catching-up between the freely flowing delayed ejecta 
and the decelerating FS. It can also be calculated analytically, if it is assumed that all the ejecta  
were released on a timescale much shorter than the observer time when the catching up occurs.
With this assumption, it can be shown that the ejecta--FS contrast Lorentz factor is constant:
\begin{equation}
 \Gamma_i/\Gamma = (1+e)^{-1/2} \cdot \left\{  \begin{array}{ll}  
                      2  &  {\rm unif\; CBM} \\  \sqrt{2} &  {\rm wind\; CBM} \end{array} \right. \;.
\label{Gi}
\end{equation}
Once the dissipated fraction is known, we track the adiabatic conversion of the internal energy into kinetic,
as described in Panaitescu \etal (1998). Although the concept of adiabatic losses implies that there will be 
a dispersion in the Lorentz factor of the swept-up CBM, we ignore it and, just as for the other two models, 
we assume that all the fluid behind the FS moves at the same Lorentz factor $\Gamma$. 

 Even if the injected energy is negligible, the RS crossing the delayed ejecta can be of relevance for the 
lower frequency afterglow emission at days after the burst. To calculate this emission, we set the typical 
electron energy behind the RS by assuming that all the dissipated energy is in the shocked ejecta, which
would be strictly correct only if the density of the shocked CBM were much larger than that of the delayed 
ejecta. Otherwise, this assumption leads to an overestimation of the post-RS electron energy. After the energy 
injection ceases, we track the evolution of the RS electron distribution, subject to adiabatic and radiative 
cooling. The tracked electron distribution is used for the calculation of the RS continuum break-frequencies. 
In contrast, for all three models, the FS electron distribution is set using the current Lorentz factor 
$\Gamma$ of the FS (see below). This last approximation is more appropriate for a uniform medium than for a 
wind, as in the former case $dM/dr \propto r^2$ while for the latter $dM/dr =$ constant.

\subsection{Model Parameters}

 As presented above, the Jet model has two dynamical parameters: the initial kinetic energy per solid angle
$\E0$ and the initial jet opening $\theta_{jet}$. The SO model has three such parameters: the energy per
solid angle $\E0$ on the axis (and in the uniform core), the core angular size $\theta_{core}$, and the 
structural parameter $q$ for the outflow envelope. The EI model has three dynamical parameters: the total 
injected energy $\Ei$, the temporal index $e$ of the injection law, and the time $t_{off}$ when the injection 
episode ends, the initial $\E0$ being less relevant if $\E0 \ll \Ei$ . In addition, all models
have another parameter which determines the outflow dynamics: the external medium density $n$ (or the wind 
parameter $A_*$). The observer location $\theta_{obs}$ relative to the outflow symmetry axis is relevant 
only for the SO model, but not for the Jet model, as long as $\theta_{obs} < \theta_{jet}$, or for the 
EI model, where the outflow is spherically symmetric.

 The calculation of the synchrotron and inverse Compton emissions (and of the radiative losses), requires 
three more parameters: one for the magnetic field strength, parameterized by the fraction $\epsB$ of the 
post-shock energy in it: 
\begin{equation}
  B^2/(8\pi) = n\, (4\Gamma + 3) (\Gamma - 1)\, m_p c^2 \cdot \epsB \;, 
\label{B}
\end{equation}
$n$ being the co-moving frame particle density in the unshocked fluid,
$m_p$ the proton mass, one for the minimal electron energy behind the shock, parameterized by the 
fraction $\epsi$ of the post-shock energy in electrons if all had the same energy $\gamma_i\, m_e c^2$:
\begin{equation}
  \gamma_i\, m_e = (\Gamma - 1) m_p \cdot \epsi \;,
\label{gammai}
\end{equation} 
$\gamma_i$ being the electron random Lorentz factor and $m_e$ the electron mass, and the index $p$ of the 
electron power-law distribution with energy:
\begin{equation}
  dN/d\epsilon \propto \epsilon^{-p} \;.
\label{p}
\end{equation}
Equations (\ref{B}) and (\ref{gammai}) apply also to the RS if $\Gamma$ is replaced by the Lorentz factor
of the shocked ejecta as measured in the frame of the incoming (unshocked) ejecta.
The parameters $\epsi$ and $\epsB$ determine the synchrotron characteristic frequency $\nu_i$ corresponding 
to the typical electron energy, as well as the self-absorption and cooling frequencies.
 
 We assume that the three microphysical parameters $\epsB$, $\epsi$, and $p$, have the same value behind
both the forward and reverse shock.

 Summarizing, the Jet model has six free parameters, the EI model seven, and the SO model eight. 
The $V$-band extinction due to dust in the host galaxy, $A_V$, which affects the observed slope of the optical 
SED and the overall optical flux, is an extra parameter for all models. To determine the host frame $A_V$
extinction, we assume an SMC-like reddening curve\footnotemark for the host galaxy and use the parameters 
inferred by Pei (1992) for SMC. 
\footnotetext{ In principle, if a simple power-law $A_\nu \propto \nu^{\kappa}$ reddening curve is assumed,
 then numerical fits to the afterglow optical continuum and the $X$-ray flux could constrain both $A_V$ and 
 the index $\kappa$ }

  All these parameters are constrained by the multiwavelength afterglow data in the following way. 
The pre-break decay index $\alpha_1$ of the optical light-curve determines the electron distribution index $p$
for the Jet and SO models (\eqs [\ref{a1sph}], [\ref{a2s0}], [\ref{a2s2}]), and the energy injection index $e$ 
for the EI model (\eq [\ref{e}]).
The post-break decay index $\alpha_2$ overconstrains the electron index $p$ for the Jet model, determines 
the structural parameter $q$ for the SO model, and the electron index $p$ for the EI model\footnotemark. 
\footnotetext{ Therefore, for the SO and EI models, the double constraint on the electron index $p$ specific 
   to the Jet model is relaxed. However this does not guarantee that the fits obtained with the SO and EI 
   models will always be better than those resulting for the Jet model, as there are differences in the 
   blast-wave dynamics of these three models and specific limits on the post-break light-curve decay 
   indices which they yield. } 
The jet opening $\theta_{jet}$ for the Jet model and the observer location $\theta_{obs}$ for the SO model 
(if the observer is outside the core) or the core angular size $\theta_{core}$ (if the observer is within the 
core) are set by the epoch of the optical light-curve break (\eqs [\ref{tjs0}] and [\ref{tjs2}]). The same 
observable determines the time $t_{off}$ when the energy injection ceases in the EI model. For the SO model, 
whenever the observer is outside the outflow core, the $\theta_{core}$ is not well-constrained because it has 
a weak effect on the resulting afterglow emission. 
The outflow energy $\E0$ (or the injected one $\Ei$), the medium density $n$ (or the wind parameter $A_*$), 
and the two microphysical parameters $\epsi$ and $\epsB$ are constrained by the afterglow spectral parameters 
-- flux at the peak of the spectrum, self-absorption, injection, and cooling frequencies -- which in turn are 
constrained by the relative intensities of the radio, optical and $X$-ray emissions (\eg Wijers \& Galama 1999, 
Granot \etal 1999). That the self-absorption frequency is not always constrained by the radio observations 
accounts for part of the resulting parameter uncertainties, but note that the parameters $\E0$, $n$, $\epsB$, 
and $\epsi$ are also constrained by matching the radio emission at late times, when the blast-wave is only 
mildly relativistic, \ie after the self-similar relativistic dynamics phase employed in analytical calculations 
of afterglow parameters. For the EI model, the initial outflow energy $\E0$ is irrelevant, as the blast-wave 
kinetic energy is that injected starting at a time before the first observation. 
Finally, the host extinction $A_V$ is constrained by the ratio of the observed optical flux to the intrinsic 
optical flux which is consistent with the $X$-ray emission (for the other afterglow parameters).

\section{Results of Numerical Modelling}

 Table \ref{Tchi} lists the reduced chi-square, $\c2nu$, for the best fits obtained with the Jet, SO, and EI 
models, for the two types of circumburst media: homogeneous and wind-like. For a uniform CBM, the best fits with 
the EI model are very poor and, for almost all afterglows, they are substantially worse than those for a wind. 
This is so because, for a uniform CBM, the observed post-break optical decay index $\alpha_2$ (given by 
equation [\ref{a1sph}]) requires a higher electron index $p$ than for a wind, leading to an optical SED 
slope ($\beta_o$) and an optical--to--$X$-ray spectral index ($\beta_{ox}$) larger than those observed. Also, 
the EI model has difficulties in accommodating both the radio and $X$-ray fluxes, with the FS emission either
overestimating the observed radio flux or underestimating the $X$-ray flux.

\begin{table*}
 \caption{ Reduced chi-square $\c2nu$ of the best fits obtained for 10 GRB afterglows with three models
          (SO = structured outflow, EI = energy injection) for light-curve breaks and for two type of 
          circumburst media: uniform ($n = const$) and wind-like ($n \propto r^{-2}$) }
\label{Tchi}
\begin{tabular}{lcccccccc}
  \hline \hline
    GRB    &     $\Da$     &  $N$  &   Jet   &   Jet           &   SO    &   SO            &   EI            \\ 
           &      (1)      &  (2)  &$n=const$&$n\propto r^{-2}$&$n=const$&$n\propto r^{-2}$&$n\propto r^{-2}$\\ 
  \hline
   980519  & $0.56\pm0.37$ &   73  &   2.6   &   1.8           &   2.4   &   1.4           &   3.0           \\
   990123  & $0.48\pm0.21$ &  112  &   2.0   &   2.3           &   1.8   &   2.2           &   1.5           \\
   990510  & $1.29\pm0.10$ &  101  &  0.78   &   3.1           &   2.1   &   4.6           &   3.0           \\
   991216  & $0.40\pm0.20$ &   84  &   2.0   &   1.8           &   1.2   &   1.2           &   3.4           \\
   000301c & $1.83\pm0.18$ &  111  &   4.4   &   8.3           &   3.3   &   7.1           &   10            \\
   000926  & $0.64\pm0.13$ &  145  &   2.2   &   3.5           &   2.2   &   2.8           &   3.3           \\
   010222  & $0.88\pm0.08$ &  175  &   2.2   &   3.9           &   1.7   &   4.0           &   4.7           \\
   011211  & $1.70\pm0.25$ &   88  &   4.7   &   8.7           &   2.3   &   4.7           &   7.6           \\
   020813  & $0.54\pm0.06$ &  105  &   1.6   &   2.6           &   1.1   &   2.6           &   1.6           \\
   030226  & $1.48\pm0.07$ &  112  &   8.5   &   17            &   4.0   &   11            &   16            \\
  \hline \hline
\end{tabular}
\hspace*{20mm}
\begin{minipage}{180mm}
  (1): magnitude of optical light-curve break, defined as the increase of the exponent of the power-law flux decay \\
  (2): number of radio, optical, and $X$-ray measurements 
\end{minipage}
\end{table*}

\begin{figure}
\psfig{figure=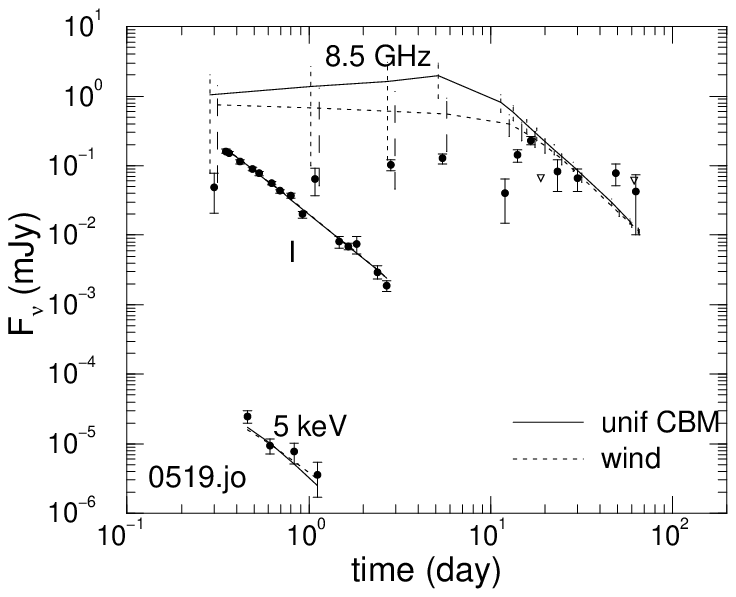,width=8.5cm,height=7cm}
\caption{ Best fit for the afterglow 980519 with the SO model. 
          The data set contains measurements at 1.4, 4.9, 8.5, 100 GHz (Frail \etal 2000b), $I, R, V, B, U$ bands 
          (Halpern \etal 1999, Vrba \etal 2000, Jaunsen \etal 2001), and 5 keV (inferred from the 2--10 keV fluxes 
          presented by Nicastro \etal 1999).
          Vertical bars on the model radio light-curve indicate the amplitude of Galactic interstellar scintillation. 
          The model $X$-ray emission is mostly inverse Compton scatterings.
          For this afterglow, a redshift was not measured. We assumed $z=1$.  }
\label{0519jo}
\end{figure}

\begin{figure}
\psfig{figure=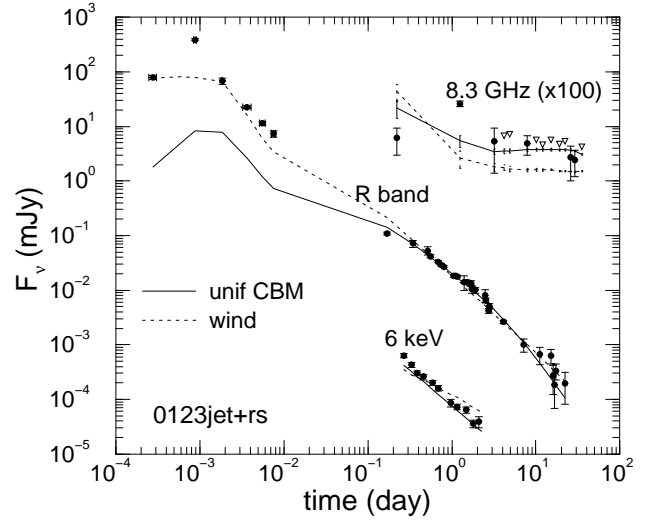,width=8.5cm,height=7cm}
\caption{ Best fit for the afterglow 990123 with the Jet model and emission from the ejecta electrons accelerated by 
          the reverse shock during the burst ($t \simeq 100$ s).
          The data set contains measurements at 1.4, 4.9, 8.3, 15, 86, 353 GHz (Galama \etal 1999, Kulkarni \etal 1999), 
          K, H, I, R, V, B (Akerlof \etal 1999, Galama \etal 1999, Kulkarni \etal 1999), and 6 keV (inferred from the
          2--10 keV fluxes reported by Piro 2000). Triangles denote $2\sigma$ observational limits on the 8.3 GHz emission.
          Radio fluxes have been shifted upwards by a factor 100, for clarity.
          For this model, the ejecta Lorentz factor $\Gamma_i$ was not constrained by equation (\ref{Gi}), because the
          assumption of instantaneous ejecta release may not be valid, and was left a free parameter. 
          The optical peak at 100 s and the radio flux before 1 day represent the reverse shock emission, which is later
          overshined by the forward shock. Model parameters are:
           $\E0 = 3 \times 10^{50} \ergsr$, $\theta_{jet} = 2.2\deg$, $n = 0.004 \cm3$, $\Ei = 1.3 \times 10^{53} \ergsr$, 
           $\Gamma_i = 5,000$, $t_{off} = 90$ s, $\epsB = 4\times 10^{-4}$, $\epsi = 0.075$, $p=2.2$, $A_V = 0.08$ for a uniform medium and
           $\E0 = 1.5 \times 10^{50} \ergsr$, $\theta_{jet} = 6.0\deg$, $A_* = 0.06$, $\Ei = 3 \times 10^{53} \ergsr$, 
           $\Gamma_i = 1,200$, $t_{off} = 170$ s, $\epsB = 7\times 10^{-5}$, $\epsi = 0.080$, $p=2.1$, $A_V = 0.07$ for a wind. 
           The addition of the early optical data has worsened the fits: $\c2nu = 14$ for the former medium, $\c2nu = 11$
           for the latter. Note that the wind model accommodates most of the early optical observations but, for either 
           type of medium, the reverse shock emission peaks in the radio too early, at about 0.1 day. }
\label{0123jet+rs} 
\end{figure}

 Tables \ref{Tjet}, \ref{Tso}, \ref{Tei}, and \ref{Tjei} list the parameters of the best fits with $\c2nu < 4$ 
for the Jet, SO, EI, and Jet+EI models, respectively, some of which are shown in Figs. \ref{0519jo}--\ref{0226jo}. 
From the parameter uncertainties determined by us (Panaitescu \& Kumar 2002) for the Jet model from the 
$\chi^2$-variation around its minimum for eight GRB afterglows, we estimate the following uncertainties for the 
best fit parameters given in Tables 2--5: 
$\sigma (\lg \E0) = \sigma (\lg \Ei) = 0.3$, $\sigma (\lg n) = 0.7$, $\sigma (\lg A_*) = 0.3$,
$\sigma (\theta_{jet}) = 0.3 \theta_{jet}$, $\sigma (\theta_{core}) = 0.2\deg$, 
$\sigma (\lg \epsB) = 1$, $\sigma (\epsi) = 0.3 \epsi$, $\sigma (p) = 0.1$, $\sigma (q) = 0.3$,
$\sigma (e) = 0.2$, $\sigma (A_V) = 0.2\, A_V$.

\begin{table*}
 \caption{ Jet model parameters for the best fits of Table 1 with $\c2nu < 4$ }
\label{Tjet}
\begin{tabular}{cccccccccccccc}
  \hline \hline
   GRB   &  $n$   & $A_*$ &  $\E0$   &$\theta_{jet}$& $\lg\epsB$ & $\lg\epsi$ &  p  & $A_V$ &     Fig       \\
         &($\cm3$)&       &($\Ten53$)&  ($\deg$)    &            &            &     &       &               \\
  \hline
  980519 &  0.1   &       &   0.8    &     2.3      &   -3.9     &    -1.2    & 2.8 &  0    & fig.2 in PK02 \\
  ...... &        &  2.0  &   0.2    &     6.7      &   -0.7     &    -1.4    & 2.4 & 0.06  &               \\
  990123 &  0.8   &       &   0.9    &     2.2      &   -3.8     &    -2.7    & 1.5 & 0.18  & \ref{0123jet} \\
  ...... &        &  0.2  &   0.7    &     2.3      &   -2.8     &    -3.1    & 1.5 & 0.16  &   .......     \\
  990510 &  0.3   &       &   0.2    &     3.1      &   -2.3     &    -1.6    & 1.8 & 0.04  & fig.3 in PK01 \\
  ...... &        &  0.4  &   1.5    &     2.1      &   -3.2     &    -1.6    & 1.8 & 0.05  &               \\
  991216 & 0.04   &       &   0.6    &     1.5      &   -1.5     &    -2.5    & 1.7 & 0.05  & \ref{1216jet} \\
  ...... &        &  0.2  &   0.8    &     2.6      &   -2.3     &    -1.9    & 1.7 & 0.05  &  ........     \\
  000926 &   22   &       &   0.1    &     8.1      &   -1.3     &    -1.1    & 2.4 & 0.14  & fig.5 in PK02 \\
  ...... &        &  1.8  &   0.2    &     5.1      &   -1.3     &    -1.1    & 2.4 & 0.12  &               \\
  010222 & 0.09   &       &   0.8    &     2.5      &   -3.1     &    -2.0    & 1.8 & 0.28  & \ref{0222jet} \\
  ...... &        &  0.1  &   0.2    &     3.8      &   -2.1     &    -2.5    & 1.5 & 0.30  &  ........     \\
  020813 & 0.07   &       &   0.7    &     2.3      &   -2.9     &    -2.1    & 1.9 & 0.25  & \ref{0813jet} \\
  ...... &        &  0.2  &   0.4    &     4.8      &   -3.5     &    -2.3    & 1.5 & 0.29  &  .......      \\
  \hline \hline
\end{tabular}
\end{table*}

\begin{table*}
 \caption{ Structured outflow parameters for the best fits of Table 1 with $\c2nu < 4$ }
\label{Tso}
\begin{tabular}{ccccccccccccccccc}
  \hline \hline
   GRB   &  $n$   & $A_*$ &  $\E0$   &  $\tc$   &  q  & $\totc$ &$\lg\epsB$&$\lg\epsi$&  p  & $A_V$ &     Fig      \\
         &($\cm3$)&       &($\Ten53$)& ($\deg$) &     &         &          &          &     &       &              \\
  \hline
  980519 &   4    &       &   0.6    &   0.9    & 1.4 &   0.9   &   -3.4   &   -1.5   & 2.7 &  0.07 & \ref{0519jo} \\
  ...... &        &  0.6  &   1.0    &   1.0    & 2.3 &   1.6   &   -3.8   &   -1.4   & 2.6 &  0.07 &   ........   \\
  990123 &   1    &       &   0.5    &   0.7    & 1.7 &   0.7   &   -3.2   &   -2.2   & 1.6 &  0.15 &              \\
  ...... &        &  0.1  &   0.2    &   1.1    & 1.5 &   1.3   &   -3.0   &   -2.6   & 1.5 &  0.14 &              \\
  990510 &   2    &       &   0.7    &   0.7    & 1.8 &   3.0   &   -2.7   &   -1.4   & 1.8 &   0   &              \\
  991216 &  0.7   &       &    3     &   0.5    & 1.6 &   1.5   &   -3.9   &   -2.0   & 1.6 &   0   &              \\
  ...... &        &  0.4  &    3     &   0.5    & 1.2 &   0.5   &   -3.8   &   -1.7   & 1.7 &  0.05 &              \\
  000301c&  0.2   &       &   2.5    &   0.8    & 2.3 &   3.3   &   -2.6   &   -1.6   & 2.4 &   0   & \ref{0301jo} \\
  000926 &   3    &       &    2     &   0.7    & 2.6 &   3.4   &   -2.8   &   -1.3   & 2.4 &  0.14 & \ref{0926jo} \\
  ...... &        &  0.5  &    2     &   0.9    & 2.6 &   1.9   &   -2.7   &   -1.5   & 2.6 &  0.18 &   ........   \\
  010222 &  0.09  &       &   0.6    &   1.7    & 2.1 &   0.7   &   -3.7   &   -1.9   & 1.8 &  0.27 &              \\
  ...... &        &   5   &   0.9    &   1.6    & 1.9 &   2.4   &   -3.9   &   -1.5   & 1.9 &  0.11 &              \\
  011211 &  0.9   &       &   0.5    &   1.1    & 2.2 &   2.6   &   -3.3   &   -1.3   & 2.3 &   0   & \ref{1211jo} \\
  020813 &  0.06  &       &   1.3    &   0.8    & 1.9 &   1.4   &   -3.4   &   -2.0   & 1.9 &  0.20 &              \\
  ...... &        &  0.06 &   0.11   &   1.1    & 1.0 &   2.0   &   -2.6   &   -2.0   & 1.6 &  0.20 &              \\  
  030226 &   1    &       &    2     &   0.5    & 2.8 &   3.6   &   -3.6   &   -1.5   & 2.4 &  0.03 & \ref{0226jo} \\
  \hline \hline
\end{tabular}
\end{table*}

\begin{table*}
 \caption{ Best fit parameters obtained with the Energy Injection model}
\label{Tei}
\begin{tabular}{cccccccccccccc}
  \hline \hline
   GRB   &  $n$   & $A_*$ &  $\Ei$   &  e  & $t_{off}$ & $\lg\epsB$ & $\lg\epsi$ &  p  & $A_V$ &     Fig        \\
         &($\cm3$)&       &($\Ten53$)&     &  (day)    &            &            &     &       &                \\
  \hline
  990123 &        & 0.003 &   0.08   & 0.7 &   0.5     &   -2.1     &    -1.0    & 2.3 & 0.04  & \ref{0123eis2} \\
  020813 &        & 0.13  &   0.7    & 0.3 &   0.9     &   -3.2     &    -2.5    & 1.8 & 0.22  & \ref{0813eis2} \\
  \hline \hline
\end{tabular}
\end{table*}

\begin{table*}
 \caption{ Best fit parameters obtained with the Jet+EI model}
\label{Tjei}
\begin{tabular}{cccccccccccccccccc}
  \hline \hline
   GRB   & $\c2nu$ &  $n$    & $A_*$ &   $\E0$  &  $\tj$ &$\Ei/\E0$&$t_{off}$&$\lg\epsB$&$\lg\epsi$&  p  & $A_V$ &     Fig        \\
         &         &($\cm3$) &       &($\Ten53$)&($\deg$)&         &  (day)  &          &          &     &       &                \\
  \hline 
  990123 &   1.9   &$10^{-3}$&       &   0.9    &   2.0  &   0.1   &   0.7    &  -3.0   &   -1.2   & 2.3 &  0    &\ref{0123jet+ei}\\
  ...... &   2.5   &         & 0.04  &   0.7    &   6.0  &   0.2   &   0.7    &  -3.5   &   -1.3   & 1.8 & 0.06  &  .......       \\
  991216 &   2.2   &$10^{-3}$&       &   1.4    &   2.1  &   0.1   &   1.5    &  -2.4   &   -1.1   & 2.4 &  0    &\ref{1216jet+ei}\\
  ...... &   1.6   &         & 0.04  &   0.8    &   3.6  &   0.2   &   1.8    &  -2.0   &   -1.5   & 2.0 &  0    &  ......        \\  
  010222 &   1.6   &$10^{-4}$&       &   0.7    &   2.0  &  0.06   &   2.5    &  -1.8   &   -1.3   & 2.3 & 0.19  &\ref{0222jet+ei}\\
  ...... &   3.6   &         & 0.06  &   0.4    &   5.0  &   0.6   &   1.7    &  -3.0   &   -1.7   & 1.7 & 0.19  &  ......        \\
  \hline \hline
\end{tabular}
\end{table*}

 Figs. \ref{0519jo}--\ref{0226jo} display the best fits obtained with the Jet model for the afterglows 010222, 
011211, 020813, and 030226, which we did not present previously, and comparable or better fits obtained with 
the SO and EI models. With the exception of the Jet model and a uniform CBM for the afterglow 990510, all best 
fits presented are not acceptable in a statistical sense, as they have $\c2nu > 1$. In general, fits with 
$\c2nu < 4$ appear adequate upon visual inspection, a significant fraction of the $\chi^2$ arising often 
from optical measurements (which have the smallest uncertainties), indicating either a small, short-lived 
afterglow fluctuation or, perhaps, an occasional outlier resulting from underestimated observational errors. 
In some cases there are also systematic discrepancies between the model fluxes and observations, most often 
in the radio. As a general rule, if the fits are poorer than $\chi^2 > 4$ for both types of circumburst media 
then the fits are not shown, with the exception of those obtained with the Jet and SO models for the afterglows 
011211 and 030226, for which we want to show that the latter model is much better than the former in
accommodating the 1 day, sharp optical light-curve break. Also as a general rule, if the fits obtained
with the Jet and SO models are comparable, only one is shown in the figures.
  
 For clarity, the figures contain only the radio optical frequencies where observations span the longest 
time, but the fitted data set included few or several other bands in these two domains (see figure captions). 
With the exception of the afterglow 000926, the available $X$-ray measurements are in only one band, which 
we use to infer the $X$-ray flux at the mid energy of that band. The amplitude of interstellar scintillation,
whose calculation is based on the treatment and maps given by Walker (1998),
is indicated with vertical bars at the time of the radio observations. In all figures, the best fit for a 
uniform CBM is shown with continuous lines while that for a wind medium with dotted lines.

\subsection{980519}

  The best fit obtained with the Jet model is discussed in Panaitescu \& Kumar (2002). The addition
  of more radio data and some optical outliers (which were previously excluded) lead now to a worse
  fit ($\chi^2$-wise) but to the same afterglow parameters, within their uncertainty.
  The best fit for the SO model, shown in Fig. \ref{0519jo}, has a $\c2nu$ comparable to that for
  the Jet model (Table 1) but is qualitatively poorer, as it overestimates the observed early radio.
  Although the measured radio fluxes are within the amplitude of the fluctuations caused by the
  interstellar Galactic gas, such a constant offset cannot be caused by scintillation (unfortunately, 
  $\chi^2$-statistics does not reflect the smaller probability of systematic discrepancies). 

\subsection{990123}

  The best fit obtained with the Jet model and a uniform medium is presented in Panaitescu \& Kumar (2001).
  There we assumed that the radio flare seen at 1 day (Kulkarni \etal 1999) arises in the GRB ejecta 
  energized by the RS, and included in the modelling only the radio measurements at 3--30 days, when the 
  8 GHz flux of this afterglow is less than $40\microJy$. This sets an upper limit on the FS peak flux,
  $F_p$, which, together with that $F_p \propto n^{1/2}$ before the jet-break and $F_p \propto n^{1/6}$ 
  after that, requires a small CBM density: $n < 10^{-2} \cm3$.

  Including the radio measurements before a few days, as well as the millimeter data, the $K$-band fluxes
  after 10 days (with which the $K$-band light-curve appears to be a single power-law, unlike the $R$-band
  emission, which exhibits a steepening at $\sim 2$ days), and other optical measurements previously left
  out of our modelling, we find now a poorer fit ($\c2nu = 2.4$ for a uniform CBM, $\c2nu = 5.1$ for a wind),
  but with the same afterglow parameters as before.

  We have also tested the ability of the Jet model including the emission from the RS crossing the GRB
  ejecta to explain the optical flash observed at 20--600 seconds (Akerlof \etal 1999), which peaked 50 s 
  after the burst at about 0.8 Jy, and the 1 day radio flare, as proposed by Sari \& Piran (1999). 
  Given that the observer injection timescale ($\sim 100$ s) is comparable to the burst duration, the
  approximation of instantaneous ejecta release and its consequence given in equation (\ref{Gi}) may be 
  invalid. For this reason, we let the ejecta Lorentz factor, $\Gamma_i$, to be a free parameter of the fit.
  Together with the injected energy, $\Ei$, it determines the number of the radiating electrons in the
  GRB ejecta and the Lorentz factor of the RS, \ie the peak flux and frequency of the synchrotron RS emission.

  The best fits obtained with the Jet model interacting with a tenuous medium (so that the FS emission 
  accommodates the late radio measurements) and emission from the GRB ejecta are shown in Fig. \ref{0123jet+rs}.
  We find that, for the same microphysical parameters $\epsi$ and $\epsB$ behind both shocks, the peak RS 
  optical emission is dimmer than observed by a factor 50 for a uniform CBM and a factor 5 for a wind,
  the latter yielding a good fit to the remainder of the early optical measurements (but not as good as
  a uniform medium to the 1 day optical break). We also find that, under the same assumption of equal
  microphysical parameters, the RS emission peaks at 8 GHz at 0.1 day, \ie a factor 10 too early. 
  In other words, taking into account the cooling of the RS electrons accelerated at $\sim 100$ s and
  the decrease on the synchrotron self-absorption optical thickness of the RS at radio frequencies, we 
  do not find a set of afterglow parameters for which the RS emission can peak in the radio as late a 1 
  day while, for the same microphysical parameters, the FS emission can accommodate the rest of afterglow 
  observations. To explain the optical flash of the afterglow 990123, the GRB ejecta must have a larger
  parameter $\epsB$ than for the forward shock, indicating that the ejecta is magnetized (Zhang, Kobayashi
  \& M\'esz\'aros 2003, Panaitescu \& Kumar 2004), a feature that may be required for by the early optical
  emission of the afterglow 021211 as well (Kumar \& Panaitescu 2003, Zhang \etal 2003).
   
  However, for the low density Jet model, an injection of ejecta in the FS at about the same time when the 
  radio flare is seen allows this model to accommodate that flare, as shown in Fig. \ref{0123jet+ei}.
  Note that the injected energy is less than that required for the FS to explain the rest of the observations, 
  \ie the incoming ejecta provide only the fresh electrons to radiate at radio frequencies, but do not alter 
  the dynamics of the FS. 

  After including the 1 day radio flare measurements, as well as the millimetre measurements at 1--10 days,
  we also find with the Jet model a higher density solution (Fig. \ref{0123jet}), $n \siml 1\cm3$, which
  is close to that obtained for other afterglows. This higher density\footnotemark is required by 
  the larger FS peak flux necessary to accommodate the 1 day flare, when $F_{8 GHz}= 0.36$ mJy. For a uniform 
  CBM, the high density fit is slightly better than the low density solution, the improvement being more 
  substantial for a wind. Note, however, that the former underestimates the radio measurements at 30 days. 
  \footnotetext{ It also lowers the cooling frequency, $\nu_c$, to just slightly above the (blueward of)
                 optical domain, which requires a hard electron distribution with index $p \simg 2\beta_{ox} 
                 \simeq 4/3$. Furthermore, a low electron parameter $\epsi$ is required for the FS peak 
                 frequency to reach 10 GHz as early as 1 day after the burst, and to explain the radio flare}

  The SO model yields fits of the same quality as the Jet model (Table 1). The best fit structural parameter
  satisfies $q > \tilde{q}(p)$ for a uniform medium and $q < \tilde{q}(p)$ for a wind (Table \ref{Tso}), thus 
  the post-break afterglow emission arises mostly from the outflow core in the former case and from the envelope
  for the latter. 
  
  The best fit obtained with the EI model for a wind medium is shown in Fig. \ref{0123eis2}. $\chi^2$-wise, 
  this low density model provides a better fit than the Jet and SO models, although it overestimates the 
  early radio measurements. Also, this solution has an extremely low wind density (Table \ref{Tei}), about 
  100 times smaller than known for Galactic WR stars. For such a tenuous wind, equation (\ref{tjs2}) implies 
  that, if the outflow opening $\theta_{jet}$ is wider than $2\deg$, then the jet-break would appear later 
  than about 30 days (\ie after the latest measurement), and the corresponding outflow energy would be larger 
  than $2\times 10^{50}$ ergs, which is less than that of the jet shown in Fig. \ref{0123jet}.
  The best fit obtained for a uniform medium is very poor (it either overestimates the radio emission or
  underproduces $X$-rays) and is not shown. 

\subsection{990510} 

  The best with the Jet model for a uniform CBM is shown in Panaitescu \& Kumar (2001). 
  A wind provides a poorer fit (Table 1), as it fails to accommodate the strong break of the optical 
  light-curves of this afterglow. Structured outflows yield poorer fits for either type of medium, the best 
  fit structural parameter being $q=1.8$ for a uniform CBM and $q=1.4$ for a wind.

\subsection{991216} 

  The best fit obtained with the Jet model and a uniform CBM is shown in Panaitescu \& Kumar (2001).
  Because the radio decay of this afterglow, $F_{8\,GHz} \propto t^{-0.77\pm0.06}$, is significantly slower 
  than that measured at optical wavelengths after the break, $F_o \propto t^{-1.65\pm0.12}$, in our previous
  modelling we have employed a double power-law electron distribution, harder (smaller index $p$) at low
  energies, to accommodate the shallower radio decay, and a softer (larger index $p$) at high energies, 
  to explain the post-break optical light-curve decay. 

  Here we use a single power-law electron distribution, to assess the ability of each model to explain all 
  the data without recourse to a light-curve decay steepening originating in a break in the electron
  distribution. Consequently, the Jet model (as well as any other single emission component model) will
  not accommodate the decays of both the radio and optical emissions of the afterglow 991216, as illustrated 
  by the higher $\c2nu$ given in Table 1 and the best fit shown in Fig. \ref{1216jet}.

  The addition of emission from the RS which energizes the ejecta catching up with the FS at about 1 day
  improves the radio fit, as shown in Fig. \ref{1216jet+ei}. A slower decay radio light-curve is obtained 
  as the FS emission overtakes that from the RS at about 10 days. As for the afterglow 990123 (Fig. 
  \ref{0123jet+ei}), the energy injected is small enough that it does not alter the dynamics of the FS. 
  However, the afterglow parameters of the Jet+EI model for 991216 are different than those obtained with 
  the Jet model (Fig. \ref{1216jet}) because we require now that the FS peaks in the radio at a later time 
  and at a lower flux ($\sim 0.1$ mJy)\footnotemark.
  \footnotetext{This requires a tenuous CBM, which increases the cooling frequency ($\nu_c \propto n^{-1}$ 
                before the jet-break and $\nu_c \propto n^{-5/6}$ after that). Going from the Jet fit to 
                the Jet+EI fit increases the cooling frequency from about $10^{15}$ Hz to $3\times 10^{17}$ Hz. 
                Then, to accommodate the optical--to--$X$-ray SED slope, the Jet+EI model requires a softer
                electron distribution than the Jet model, which in turn yields a faster post-break decay of 
                the afterglow light-curve, as shown in Fig. \ref{1216jet+ei}, and a poorer fit to the optical 
                emission after 10 days}

  The SO model yields a best fit which is very similar to that shown in Fig. \ref{1216jet} for the Jet model. 
  The smaller intensity-averaged source size resulting in the SO model leads to a larger amplitude for the
  interstellar scintillation and, thus, to a better $\chi^2$-wise fit (Table 1), but this model also fails 
  to accommodate the slow radio decay. The best fit structural parameter satisfies $q \siml \tilde{q}(p)$
  (Table \ref{Tso}), \ie the outflow envelope emission dominates over that from the core after the optical break.
  
\subsection{000301c} 

  The best fit for the Jet model is discussed in Panaitescu (2001). As for the afterglow 991216, the slower
  radio decay of 000301c, $F_{8\,GHz} \propto t^{-1.1\pm0.3}$, and its faster optical fall-off, $F_o \propto 
  t^{-2.83\pm0.12}$, prompted us to consider a double power-law electron distribution. Furthermore, the 
  large break magnitude $\Da = 1.8 \pm 0.2$ exhibited by the optical light-curve 000301c (the largest in the
  entire sample - Table 1) exceeds that allowed by the Jet model for the electron distribution index $p$ 
  required by the pre-break decay index and optical SED slope (Panaitescu 2005). Such a strong break is also 
  better explained if there is a contribution from the passage of a spectral break.
  However, since we want to test the three models for light-curve break without any contribution from another
  mechanism, we list in Table 1 the best fit obtained previously with a soft electron distribution.

  Fig. \ref{0301jo} shows the best fit obtained with the SO model, which is only slightly better than that
  of the Jet model, has similar parameters, and shares the same deficiencies: it fails to explain the sharpness
  of the optical light-curve break and yields a radio emission decaying faster than observed. We note that
  optical measurements between 3.0 and 4.5 days, when a light-curve bump is seen, have been excluded from the fit.

\subsection{000926}

  The best fit obtained with the Jet model is shown in Panaitescu \& Kumar (2002). After including in the
  data set the near-infrared measurements, the noisy $I$-band, and some optical outliers which were previously 
  excluded, the best fit is $\chi^2$-wise poorer but the fit parameters are the same as before (within their 
  uncertainty). An equally good fit can be obtained with the SO model (Fig. \ref{0926jo}). We note that for
  both the Jet and S0 models, the $X$-ray emission is mostly inverse Compton scatterings.

\subsection{010222} 

  The best fit obtained with the Jet model is shown in Fig. \ref{0222jet}. Note that the model radio 
  light-curve decays after 1 day faster than observed. For a uniform CBM, the best fit obtained with a 
  structured outflow is slightly better in the radio, but it too fails to explain the slow radio decay. 
  For a wind, the best fit obtained with the SO model overestimates the millimeter emission 
  [$F_{250\,GHz}(1-100\,d) \simeq 1.2$ mJy, $F_{350\,GHz}(0.3-20\,d) \simeq 3.9$ mJy] attributed to the 
  host galaxy by Frail \etal (2002), who also suggest a possible host synchrotron emission at 8.5 GHz 
  of $\sim 20\, \microJy$. This is only marginally consistent with the $F_{1.4\,GHz} (447\,d) = -1 \pm 
  35 \microJy$ reported by Galama \etal (2003) and the host synchrotron SED, $F_\nu \propto \nu^{-0.75}$, 
  adopted by Frail \etal (2002), which imply a host flux $F_{8\, GHz} = 0 \pm 9\, \microJy$. 

  If we subtract a radio host contribution of $F_\nu^{(host)} = 20\, (\nu/8\,{\rm GHz})^{-0.75} \,\microJy$ 
  from the 44 radio measurements, then, within the Jet model, the contribution to $\chi^2$ of the radio data 
  decreases from 107 to 72 for a uniform medium, and from 103 to 53 for a wind, with similar changes for the 
  SO model. Though these are significant improvements, the best fit still underestimates the radio flux 
  measured after 10 days, because the optical measurements have a smaller uncertainty and determine the 
  electron index $p$ and, implicitly, the decay of the model radio emission. In general, one-component models 
  cannot accommodate both the host-subtracted radio emission, $F_{8\,GHz} \propto t^{-0.76\pm0.12}$, of 
  the afterglow 010222 at 1--200 days, and the optical decay, $F_o \propto t^{-1.78\pm0.08}$, observed at 
  10--100 days.

  A better fit to the radio emission of the afterglow 010222 may be obtained with the Jet model if, in 
  addition to the FS emission, there is a RS contribution to the early radio afterglow, as illustrated in 
  Fig.  \ref{0222jet+ei} for a uniform CBM. Just as for the afterglows 990123 and 991216, the incoming 
  ejecta carry less energy than that in the FS (and do not alter its dynamics) and the medium density is 
  lower than for the Jet model because the FS peak flux must be smaller, to match the radio flux measured 
  after 10 days, when the FS synchrotron peak frequency crosses the radio domain. 

\subsection{011211} 

  As illustrated in Fig. \ref{1211jet}, even for a uniform CBM, the Jet model has difficulty in accommodating 
  the sharp break exhibited by the optical light-curve of this afterglow, assuming that the 1--2 day optical 
  emission is not a fluctuation. Such fluctuations have been seen in the afterglows 000301c, 021004, and 030329, 
  but the lack of a continuous monitoring for 011211 prevents us to determine if its optical emission at 1 day 
  is indeed a fluctuation.
  A stronger break can be obtained with the SO model when the brighter, outflow core becomes visible to an 
  observer located outside it, hence the better fit that the SO model yields (Fig. \ref{1211jo}).

\subsection{020813} 

  The best fit obtained with the Jet model is shown in Fig. \ref{0813jet}. Structured outflows yield fits with 
  $q > \tilde{q}(p)$ for a uniform CBM and $q < \tilde{q}(p)$ for a wind, \ie with the post-break emission
  arising mostly in the outflow core and envelope, respectively. A significant improvement over the Jet model 
  is obtained only for a uniform CBM (Table 1). 

  A fit of comparable quality is also obtained with the EI model for a wind medium (Fig. \ref{0813eis2}), 
  but not for a uniform CBM, which overproduces radio emission and largely underestimates the observed $X$-ray 
  fluxes. For the best fit parameters given in  Table \ref{Tei}, equation (\ref{tjs2}) shows that a jet-break 
  would occur after the last measurement (40 days) if the outflow opening were larger than $5\deg$, which 
  implies a minimum collimated energy of $2\times 10^{51}$ ergs, which is 3 times larger than that of the 
  jet shown in Fig. \ref{0813jet}. Therefore, just as for 990123, the EI model may not involve significantly 
  larger energies than the Jet model, if allowance is made for a possible outflow collimation.

\subsection{030226} 

  As for 011211, the strong optical light-curve break of this afterglow is better accommodated by the SO model 
 than the Jet model (Fig. \ref{0226jet} and \ref{0226jo}), though an acceptable fit to the optical break is
 obtained only for a uniform CBM (in addition, for a wind medium, the SO model overproduces radio emission). 
 However, given the poor coverage before 0.3 days, it is possible that the measured optical emission at that 
 time is a fluctuation below that expected in the Jet model. 

\subsection{970508}

 Although the decay of the optical emission of this afterglow does not exhibit a steepening, being so far the
 longest monitored power-law fall-off (2--300 days), it can be explained with the Jet model if its unusual
 rise at 1--2 days is interpreted as a jet becoming visible to an observer located outside the initial
 jet opening (see fig. 1 in Panaitescu \& Kumar 2002). This requires a rather wide jet, with $\theta_{jet} = 
 18\deg$, and yields a poor fit to the $X$-ray emission ($\c2nu = 2.8$). A less energetic envelope surrounding
 this jet would produce the prompt $\gamma$-ray emission and would account for the flat optical emission
 prior to the 2 day optical rise, hence this model is similar to a Structured Outflow with a wide uniform
 core and an envelope whose structure is poorly constrained by observations.

 An other possible reason for the 2 day sharp rise of this afterglow is a powerful energy injection episode
 in which the incoming ejecta carry an energy larger than that already existing in the FS (Panaitescu, M\'esz\'aros
 \& Rees 1998), \ie the blast-wave dynamics is altered drastically by the energy injection. Fig. \ref{0508ei}
 shows the best with the EI model. Note that the given that the light-curve rise is achromatic, as it arises
 from the blast-wave dynamics, which explains why the $X$-ray light-curve cannot be accommodated.

\section{Conclusions}

 In this work we expand our previous investigation (Panaitescu \& Kumar 2003) of the {\bf Jet model} to four 
other GRB afterglows for which an optical light-curve break was observed: 010222 (for which radio observations 
were not available at the time of our previous modelling), 011211, 020813, and 030226. We also note that, here, 
we have not employed a broken power-law electron distribution to explain both the slower decaying radio 
emission and the faster falling-off optical light-curve observed for some afterglows (\eg 991216 and 000301c).

 Our prior conclusion that a homogeneous circumburst medium provides a better fit to the broadband afterglow 
emission than environments with a wind-like stratification stands (Table 1). For afterglows with a sharp 
steepening $\Da$ of the optical light-curve decay (990510, 000301c, 011211, and 030226), this is 
due to that the transition between the jet asymptotic dynamical regimes (spherical expansion at early times 
and lateral spreading afterward) occurs faster for a homogeneous medium than for a wind (Kumar \& Panaitescu 
2000). For the afterglows with shallower breaks, a wind medium (as expected if the GRB progenitor is a 
massive star) provides an equally good fit as (or, at least, not much worse than) a homogeneous medium. 

 For the acceptable fits (Table 2) obtained with the Jet model and for a uniform medium, the jet initial kinetic 
energy, $E_{jet}$, is between 2 and 6 $\times 10^{50}$ ergs, the initial jet opening angle, $\theta_{jet}$, 
is between $2\deg$ and $3\deg$ (with 000926 an outlier at $8\deg$), and the medium density, $n$, between 0.05 
and 1 $\cm3$ (with 000926 an outlier, again, at $20\cm3$). For a wind, $E_{jet}$ has about the same range, 
$\theta_{jet}$ ranges from $2\deg$ to $6\deg$, while the wind parameter $A_*$ is between 0.1 and 2. 70\% of 
the 64 Galactic WR stars analyzed by Nugis \& Lamers (2000) have a $A_*$ parameter in this range; for the 
rest $A_* \in (2,5)$. We note that the $E_{jet}$ resulting from our numerical fits is uncertain by a factor 2, 
$\theta_{jet}$ by about 30\%, $n$ by almost one order of magnitude, and $A_*$ by a factor of 2.

 Although the inferred wind density parameter $A_*$ is in accord with the measurements for WR stars a 
uniform medium is instead favoured by the fits to the afterglow data. This is an interesting issue for the 
established origin of long bursts in the collapse of massive stars. Wijers (2001) has proposed that the 
termination shock resulting from the interaction between the wind and the circumstellar medium could homogenize 
the wind. The numerical hydrodynamical calculations of Garcia-Segura, Langer \& Mac Low (1996) show that, 
for a negligible pressure $nT$ of the circumstellar medium (about $10^4 \cm3 K$), the termination shock of 
the RSG wind has a radius of 10 pc, outside a shell of uniform density is form. However, this radius is much 
larger than the distance of about 1 pc where the afterglow emission is produced. Chevalier, Li \& Fransson 
(2004) made the case that, if the burst occurred in an intense starburst region, where the interstellar pressure 
is about $10^7 \cm3 K$, then the shocked RSG wind would form bubble of uniform density of about $1\, \cm3$ 
extending from 0.4 to 1.6 pc, in accord with the fit densities and the blast-wave radius $r = 8 c\Gamma^2 t/(1+z) =
0.5\, (E_{0,53}/n_0)^{1/4} [t_d/(1+z)]^{1/4}$ pc, resulting from equation (\ref{Gs0}). 

 The {\bf Structured Outflow model} involving an energetic, uniform core, surrounded by a power-law envelope 
(equation [\ref{Eq}]) which impedes the lateral spreading of the core, retains most of the ability of the
Jet model to yield a steeply decaying optical light-curve after the core (or its boundary) becomes visible 
to an observer located outside the core opening (or within it), while accommodating, at the same time, the 
observed slope of the optical spectral energy distribution (SED). This is so because, in both models, more 
than half of the increase $\Da$ in the light-curve power-law decay index is due to the finite opening of 
the bright(er) ejecta. For a jet, the remainder of the steepening $\Da$ is caused by the lateral spreading. 
For a structured outflow, an extra contribution to the steepening, which can lead to a $\Da$ even larger 
than that produced by a jet, results if the observer is located outside the core opening. 

 For a uniform medium, we find that the Structured Outflow model provides a better fit than the Jet model 
for 6 of the 10 afterglows analyzed here (Table 1). For a wind, the former model works better for three 
afterglows, fits of comparable quality being obtained for the other six. Only the afterglow 990510 is better 
explained with a jet, for either type of circumburst medium. Within the framework of structured outflows, 
8 out of 10 afterglows are better fit with a uniform medium than with a wind, which is partly due to that, 
just as for a jet, the light-curve steepening is sharper if the medium is uniform (Panaitescu \& Kumar 2003). 

 For the "acceptable" fits ($\c2nu < 4$) obtained with the SO model, the ejecta kinetic energy per solid angle 
in the core, $\E0$, ranges from 0.5 to 3 $\times 10^{53}\,\ergsr$ for either type of medium, 
with an uncertainty of a factor 3. The angular opening of the core, $\theta_{core}$, is between $0.5\deg$ 
and $1\deg$ for a homogeneous medium, and about $1\deg$ for a wind (991216 being an outlier at $0.5\deg$), 
with an uncertainty less than 50\%. The structural parameter $q$ (equation [\ref{Eq}]), which characterizes 
the energy distribution in the outflow envelope, is found to be between 1.5 and 2.7 for a uniform medium, 
and between 1.0 and 2.3 for a wind (with 000926 being an outlier at $2.9$). These values are consistent 
with that proposed by Rossi \etal (2002), $q = 2$, to explain the quasi-universal jet energy resulting 
in the Jet model. For structured outflows, the ejecta kinetic energy contained within an opening of twice 
the location of the observer (relative to the outflow symmetry axis), $\theta_{obs}$, ranges from 1 to 4 
$\times 10^{50}$ ergs, for either type of medium. Hence, the energy budget required by the Structured Outflow 
model is very similar to that of the Jet model. The best fit medium densities obtained with the Structured 
Outflow model are similar to those resulting for a Jet.
 
 We find that the {\bf Energy Injection model}, where a light-curve decay steepening is attributed to the 
cessation of energy injection in the forward shock, can be at work only in two afterglows, 990123 and 020813. 
The major shortcoming of the EI model is that, in order to explain the steep fall-off of the optical light-curve 
after the break with a spherical outflow, an electron distribution that is too soft (\ie an index $p$ too large) 
is required to accommodate the relative intensities of the radio and $X$-ray emissions. Low density solutions 
($n < 1 \cm3$, $A_* < 0.1$), for which the $X$-ray emission is mostly synchrotron, yield a cooling frequency 
that is too low, resulting in model $X$-ray fluxes underestimating the observations by a factor of at least 10. 
High density solutions ($n > 10 \cm3$, $A_* > 1$), for which the $X$-ray is mostly inverse Compton scatterings, 
produce a forward shock peak flux that is too large, which leads to model radio fluxes overestimating the data 
by a factor of 10.

 We note that, due to adiabatic cooling, the GRB ejecta electrons which were energized by the reverse shock 
during the burst, do not contribute significantly to the afterglow radio emission at days after the burst. 
Numerically, we find that, if the microphysical parameters are the same for both the reverse and forward shocks,
then the 1 day radio flare of the afterglow 990123 cannot be explained with the emission from the GRB ejecta 
electrons, as proposed by Sari \& Piran (1999), as the synchrotron emission from these electrons peaks in the 
radio earlier than 0.1 day, regardless of how relativistic is the reverse shock. For the reverse shock emission 
to be significant at 1 day after the burst, there must be some other ejecta catching up with the decelerating 
outflow at that time or the assumption of equal microphysical parameters for both shocks must be invalid
(Panaitescu \& Kumar 2004). Furthermore, the optical emission from the ejecta electrons is below that observed 
at 100 s (Akerlof \etal 1999) by a factor 5 for a wind and a factor 50 for a uniform medium.

\begin{figure}
\psfig{figure=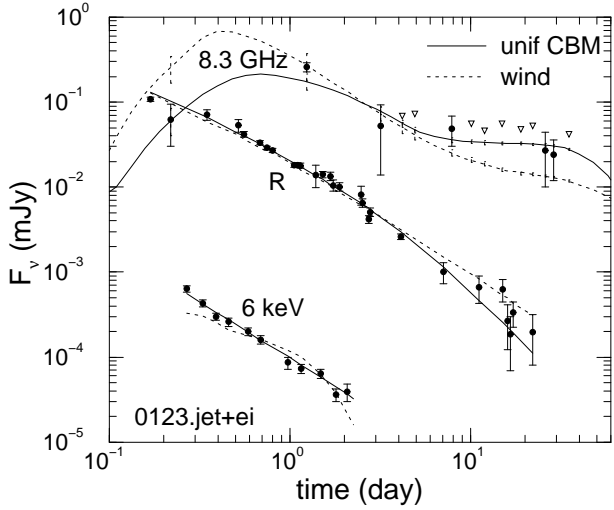,width=8.5cm,height=7cm}
\caption{ Best fit for the afterglow 990123 with the Jet+EI model for a uniform medium and a wind.
          The peak of the radio light-curve before 1 day represents the reverse shock emission, while the forward
          shock peak frequency crosses the radio domain after 10 days. }
\label{0123jet+ei}
\end{figure}

\begin{figure}
\psfig{figure=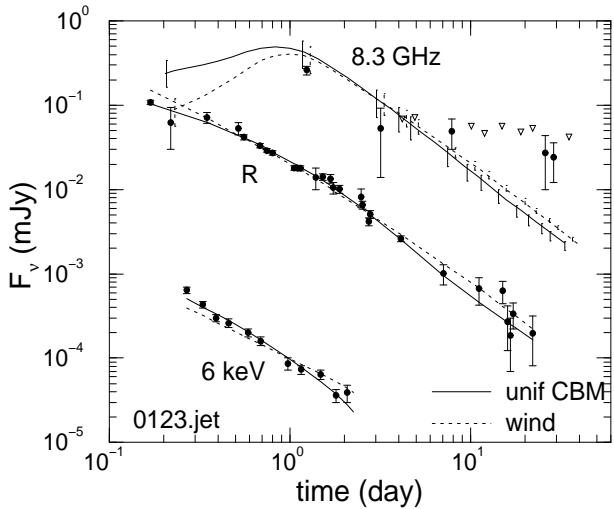,width=8.5cm,height=7cm}
\caption{ Best fit for the afterglow 990123 with the Jet model.
          Note that, in contrast to the fits shown in figure \ref{0123jet+ei}, the medium density of these fits 
          is similar to that determined for other afterglows. However, the electron energy distribution has
          unusual parameters $\epsi$ and $p$.  } 
\label{0123jet}
\end{figure}

\begin{figure}
\psfig{figure=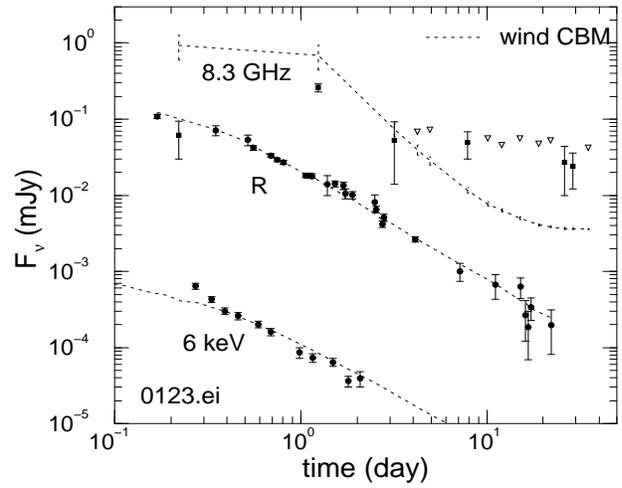,width=8.5cm,height=7cm}
\caption{ Best fit for the afterglow 990123 with the EI model for a wind medium.
         The model radio emission until 10 days arises in the reverse shock which has energized the incoming 
         ejecta, after which the forward shock emission overtakes it.  }
\label{0123eis2}
\end{figure}

\begin{figure}
\psfig{figure=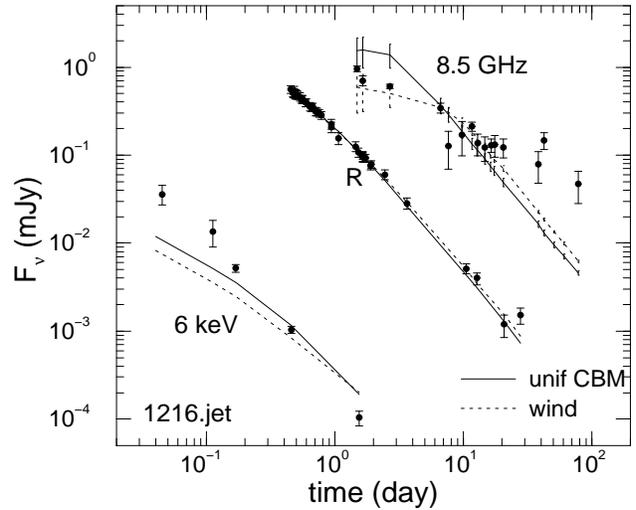,width=8.5cm,height=7cm}
\caption{ Best fit for the afterglow 991216 with the Jet model. 
     The data set consists of measurements at 1.4, 4.9, 8.5, 15 GHz (Frail \etal 2000c), $K, H, J, I, R, V$ 
     bands (Garnavich \etal 2000, Halpern \etal 2000), and 6 keV (inferred from the 2--10 keV fluxes reported 
     by Corbet \& Smith 1999, Piro \etal 1999, Takeshima \etal 1999). 
     Note that this model cannot explain the slower radio decay observed at about the same time with the faster 
     optical fall-off.}
\label{1216jet}
\end{figure}

\begin{figure}
\psfig{figure=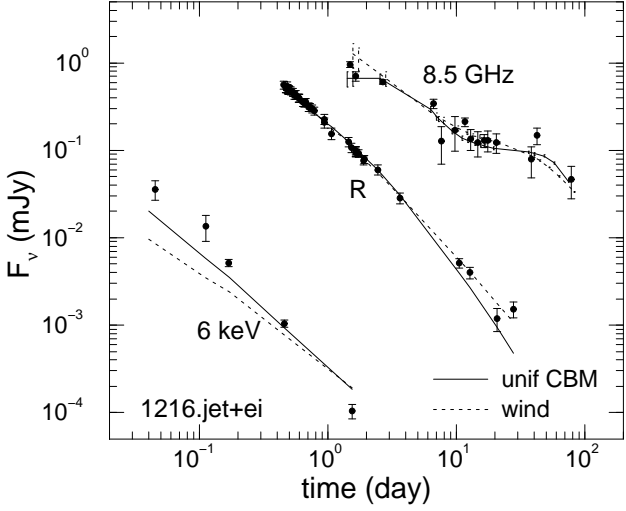,width=8.5cm,height=7cm}
\caption{ Best fit for the afterglow 991216 with the Jet+EI model for a uniform medium and a wind. 
      The addition of the reverse shock, which yields the 2 day peak of the radio emission, improves 
      the fit to the radio data compared to that obtained with the Jet model (figure \ref{1216jet}). }
\label{1216jet+ei}
\end{figure}

\begin{figure}
\psfig{figure=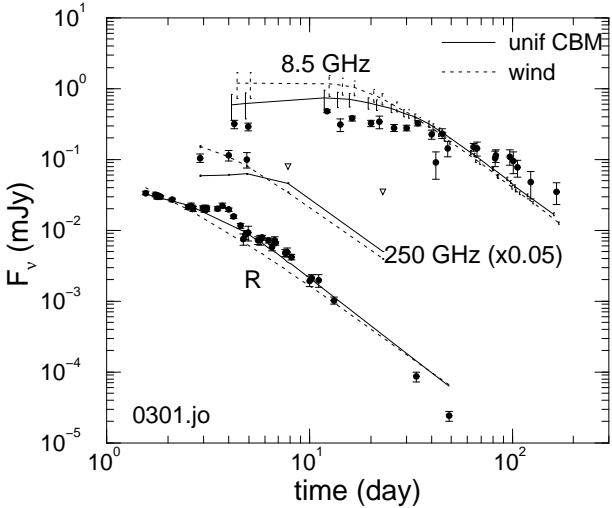,width=8.5cm,height=7cm}
\caption{ Best fit for the afterglow 000301c for the SO model. 
     The data set contains measurements at 1.4, 4.9, 8.5, 15, 22, 100, 250 350 GHz (Berger \etal 2000, Smith \etal 2001, 
     Frail \etal 2003), and $K, J, I, R, V, B, U$ bands (Jensen \etal 2001, Rhoads \& Fruchter 2001). 
     The 250 GHz fluxes and their $2\sigma$ limits (triangles) have been shifted downward by a factor 20, 
     for clarity. Measurements between 3.0 and 4.5 days, when there is a fluctuation in the optical emission,
     have been left out. 
     The best fit parameters for the uniform medium are given in Table \ref{Tso}, those for the wind medium are
              $\E0 = 1.5 \times 10^{53} \ergsr$, $\theta_{core} = 0.9\deg$, $\totc = 2.6$, $q=1.4$,
              $A_* = 0.6$, $\epsB = 3\times 10^{-3}$, $\epsi = 0.018$, $p=2.4$, $A_V = 0.06$.  }

\label{0301jo}
\end{figure}

\begin{figure}
\psfig{figure=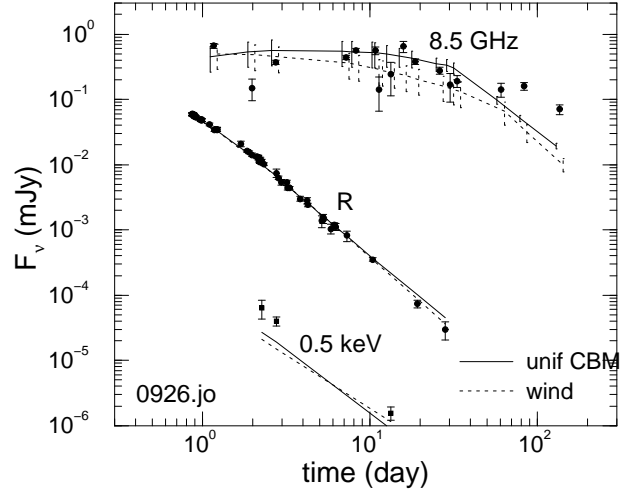,width=8.5cm,height=7cm}
\caption{ Best fit to the afterglow 000926 with the SO model. 
     The data consists of measurements at 1.4, 4.9, 8.5, 15, 22, 99 GHz (Harrison \etal 2001, Frail \etal 2003), 
     $K, H, J, I, R, V, B, U$ bands (Fynbo \etal 2001, Harrison \etal 2001, Price \etal 2001), 0.5 and 3 keV 
     (inferred from the 0.2--1.5 keV and 2--10 keV fluxes reported by Harrison \etal 2001 and Piro \etal 2001). }
\label{0926jo}
\end{figure}

\begin{figure}
\psfig{figure=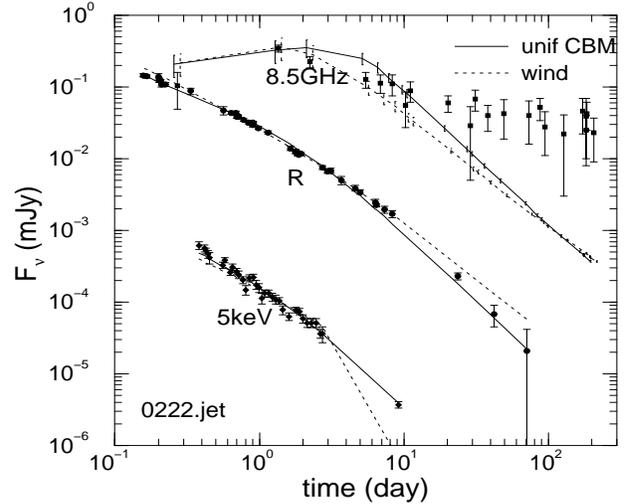,width=8.5cm,height=7cm}
\caption{ Best fit to the afterglow 010222 with the Jet model. 
     The data set contains measurements at 1.4, 4.9, 8.5, 15, 22 GHz (Galama \etal 2003), $K, J, I, R, V, B, U$ 
     bands (Masetti \etal 2001, Stanek \etal 2001), and 5 keV (inferred from the 2-10 keV fluxes reported by 
     in't Zand \etal 2001). 
     The steep $X$-ray fall-off after 2 days for a wind medium is due to a high energy cutoff of the electron 
     distribution, corresponding to a total electron energy of 50 per cent of the post-shock energy. }
\label{0222jet}
\end{figure}

\begin{figure}
\psfig{figure=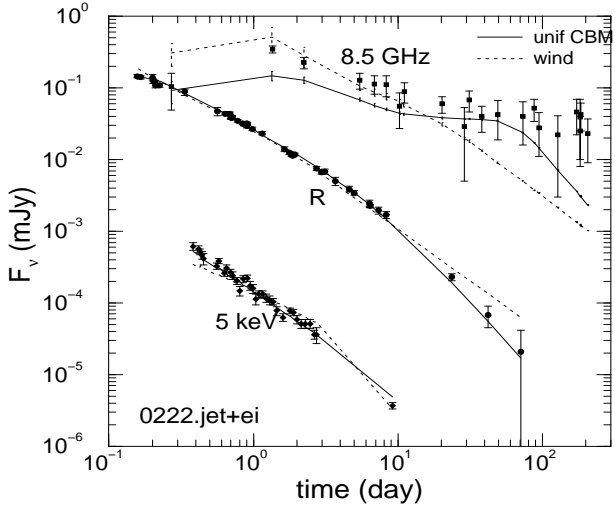,width=8.5cm,height=7cm}
\caption{ Best fit for the afterglow 010222 with the Jet+EI model.
      For a uniform medium the forward shock component accommodates better the radio data after 10 days than for a wind.
      The 1 day hump in the radio light-curve is the reverse shock emission, the forward shock overtaking at 
      10 days. }
\label{0222jet+ei}
\end{figure}

\begin{figure}
\psfig{figure=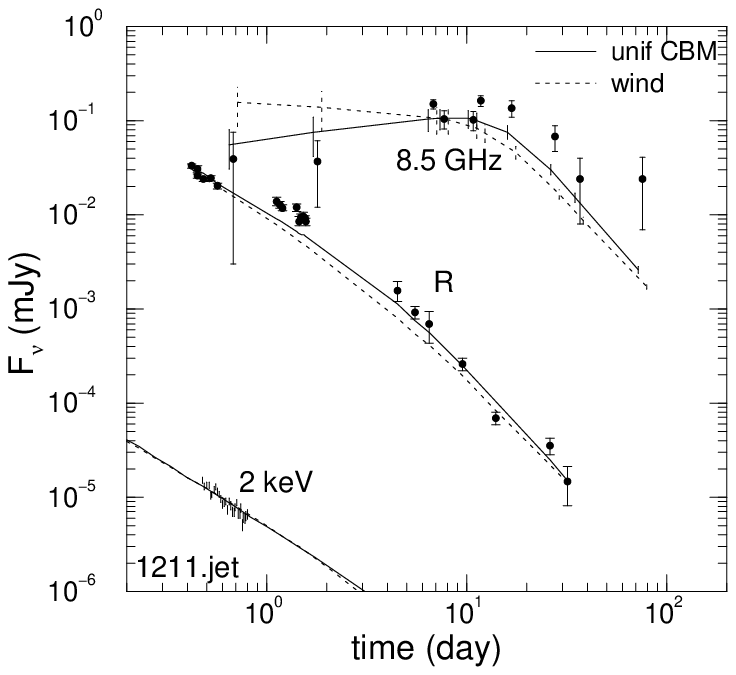,width=8.5cm,height=7cm}
\caption{ Best fit for the afterglow 011211 with the Jet model. 
      The data set contains measurements at 8.5, 22 GHz (http://www.aoc.nrao.edu/~dfrail/011211.dat), 
      $K, J, I, R, V, B, U$ bands (Holland \etal 2002, Jakobsson \etal 2003), and 2 keV (inferred from 
      the 0.2--5 keV fluxes measured by Borozdin \& Trodolyubov 2003).  
      The best fit parameters are $\E0 = 2 \times 10^{52} \ergsr$, $\theta_{jet} = 5.5\deg$, $n = 1.0 \cm3$,
                                  $\epsB = 6\times 10^{-4}$, $\epsi = 0.047$, $p=2.3$, $A_V = 0$ for a uniform medium and
                                  $\E0 = 4 \times 10^{52} \ergsr$, $\theta_{jet} = 4.0\deg$, $A_* = 0.6$,
                                  $\epsB = 5\times 10^{-4}$, $\epsi = 0.030$, $p=2.2$, $A_V = 0$ for a wind.  }
\label{1211jet}
\end{figure}

\begin{figure}
\psfig{figure=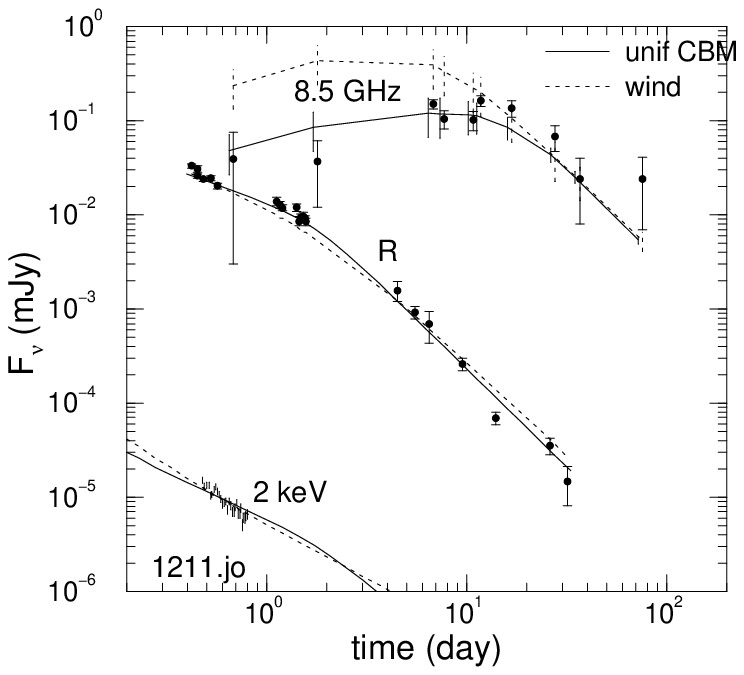,width=8.5cm,height=7cm}
\caption{ Best fit for the afterglow 011211 with the SO model. 
      The best fit parameters for the uniform medium are given in Table \ref{Tso}, those for the wind medium are
              $\E0 = 1.2 \times 10^{53} \ergsr$, $\theta_{core} = 0.9\deg$, $\totc = 3.0$, $q=2.3$,
              $A_* = 0.7$, $\epsB =  10^{-3}$, $\epsi = 0.0081$, $p=2.3$, $A_V = 0$.  }
\label{1211jo}
\end{figure}

\begin{figure}
\psfig{figure=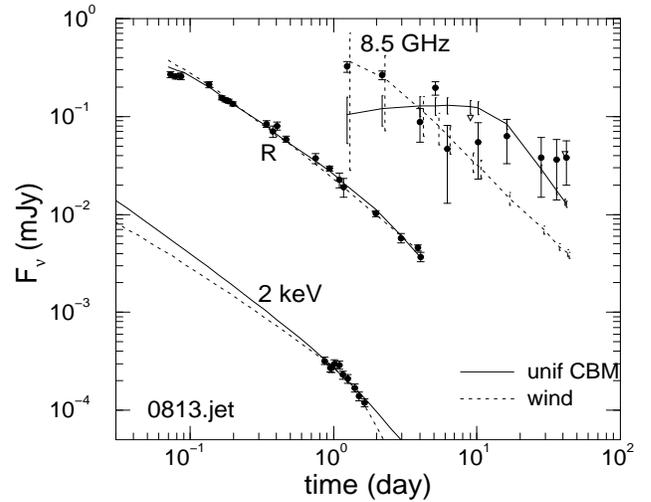,width=8.5cm,height=7cm}
\caption{ Best fit for the afterglow 020813 with the Jet model. 
      The data set contains measurements at 4.5, 8.5 (http://www.aoc.nrao.edu/~dfrail/grb020813.dat), 250 GHz 
      (Bertoldi \etal 2002), $K, H, J, I, R, V, B, U$ bands (Covino \etal 2003, Gorosabel \etal 2004), and 
      2 keV (inferred from the 0.6--6 keV fluxes measured by Butler \etal 2003).  
      For the latter, the sharp steepening of the $X$-ray light-curve at $\simg 1$ day is due to the electron 
      high-energy cut-off.  }
\label{0813jet}
\end{figure}

\begin{figure}
\psfig{figure=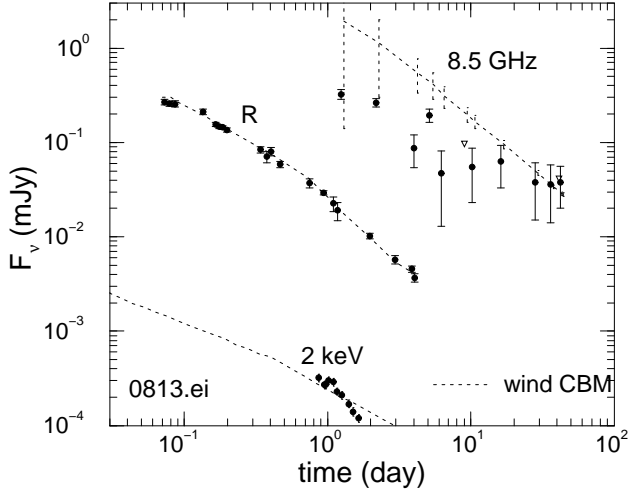,width=8.5cm,height=7cm}
\caption{ Best fit for the afterglow 020813 with the EI model for a wind medium.
      The reverse and forward shock radio emissions are equal at 1 day, after that the forward shock emission is stronger.  }
\label{0813eis2}
\end{figure}

\begin{figure}
\psfig{figure=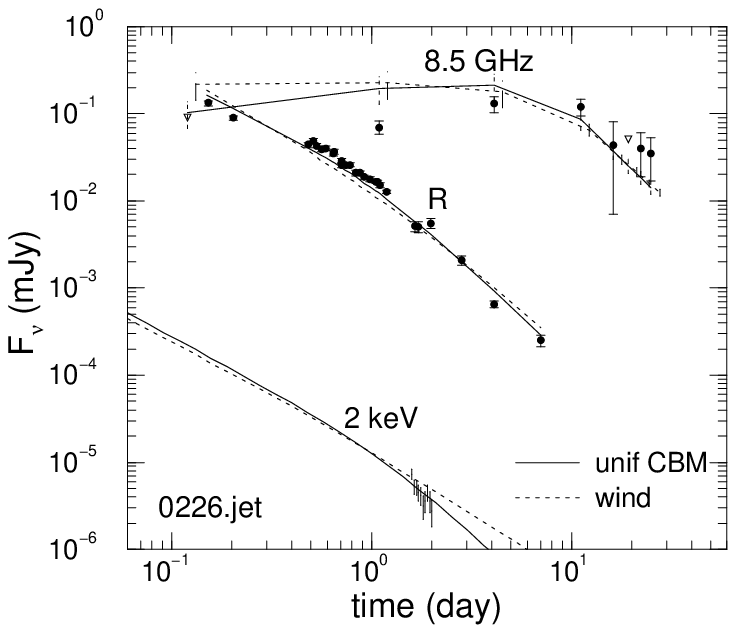,width=8.5cm,height=7cm}
\caption{ Best fit for the afterglow 030226 with the Jet model. 
      The data set contains measurements at 8.5, 15, 22 GHz (http://www.aoc.nrao.edu/~dfrail/grb030226.dat), 
      $K, H, J, I, R, V, B, U$ bands (Klose \etal 2004, Pandey \etal 2004), and 2 keV (inferred from the 
      0.3--10 keV count rate reported by Klose \etal 2004). 
      The best fit parameters are $\E0 = 6 \times 10^{52} \ergsr$, $\theta_{jet} = 2.8\deg$, $n = 1.6 \cm3$,
                                  $\epsB = 2\times 10^{-4}$, $\epsi = 0.029$, $p=2.2$, $A_V = 0.09$ for a uniform medium and
                                  $\E0 = 9 \times 10^{52} \ergsr$, $\theta_{jet} = 2.6\deg$, $A_* = 0.5$,
                                  $\epsB = 4\times 10^{-4}$, $\epsi = 0.016$, $p=2.1$, $A_V = 0.04$ for a wind.  }

\label{0226jet}
\end{figure}

\begin{figure}
\psfig{figure=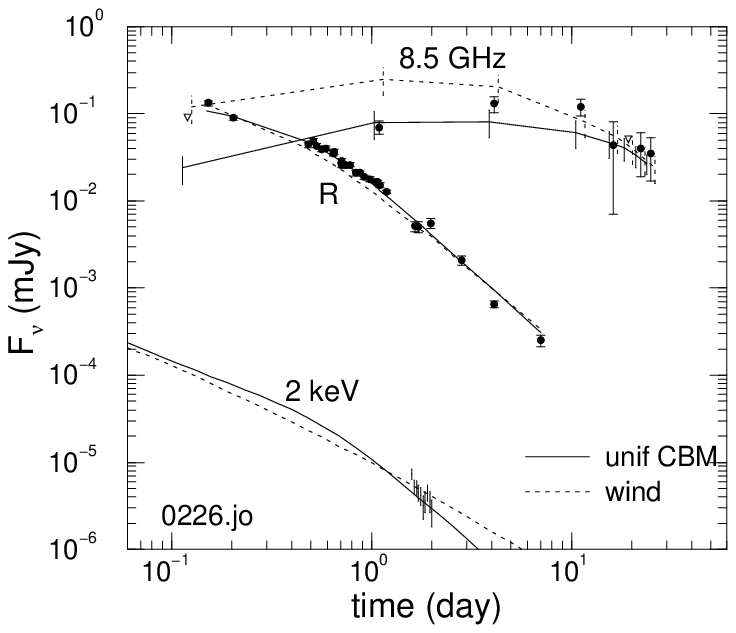,width=8.5cm,height=7cm}
\caption{ Best fit for the afterglow 030226 with the SO model.  
      The best fit parameters for the uniform medium are given in Table \ref{Tso}, those for the wind medium are
              $\E0 = 1.3 \times 10^{53} \ergsr$, $\theta_{core} = 0.5\deg$, $\totc = 4.1$, $q=2.8$,
              $A_* = 0.5$, $\epsB = 6\times 10^{-4}$, $\epsi = 0.017$, $p=2.2$, $A_V = 0.03$.  }

\label{0226jo}
\end{figure}

\begin{figure}
\psfig{figure=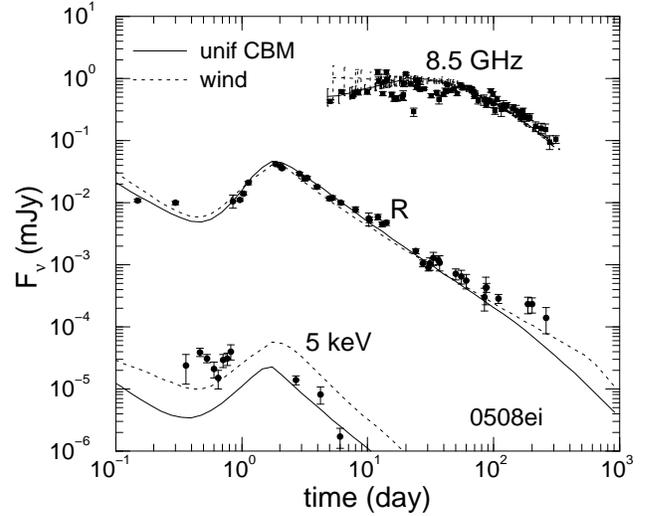,width=8.5cm,height=7cm}
\caption{ Best fit for the afterglow 970508 with the EI model.
     The data set contains 309 measurements at 1.4, 4.9 GHz (Frail, Waxman \& Kulkarni 2000a), 86 GHz (Bremer \etal 1998),
     $K, I, R, V, B$ bands (Chary \etal 1998, Galama \etal 1998, Garcia \etal 1998, Pedersen \etal 1998, Sahu \etal 1997, 
     Sokolov \etal 1998), and 5 keV (inferred from 2--10 keV fluxes and spectral slope given in Piro \etal 1998). 
     The fit parameters are: $\Ei = 9 \times 10^{51} \ergsr$, $e = 2.4$, $t_{off} = 1.6$ d, $n = 0.9 \cm3$, 
                             $\epsB = 10^{-3}$, $\epsi = 0.11$, $p=2.3$, $A_V = 0.15$ for a uniform medium ($\c2nu = 4.5$) and
                             $\Ei = 7 \times 10^{51} \ergsr$, $e = 2.7$, $t_{off} = 1.8$ d, $A_* = 0.7 \cm3$, 
                             $\epsB = 2\times 10^{-2}$, $\epsi = 0.063$, $p=2.0$, $A_V = 0.18$ for a wind ($\c2nu = 5.9$). } 
\label{0508ei}
\end{figure}

 A delayed injection of ejecta, carrying less kinetic energy than that already existing in the forward shock 
(\ie not altering the forward shock dynamics), may be at work in the afterglows 990123, 991216, and 010222, 
whose radio decay is substantially slower than that observed in the optical after the break. As shown by us 
(Panaitescu \& Kumar 2004), such a decoupling of the radio and optical emissions cannot be explained by models 
based on a single emission component (forward shock) and require the contribution of the reverse shock. Figs. 
\ref{0123jet+ei}, \ref{1216jet+ei}, and \ref{0222jet+ei}, illustrate that the sum of the reverse and forward 
shock emissions yields a slower decaying radio emission, consistent with the observations.
However, for the forward shock peak flux to be as dim as observed in the radio after 10 days, when the forward 
shock peak frequency decreases to 10 GHz, an extremely tenuous medium is required: $n \sim 10^{-3}\cm3$ or 
$A_* \sim 0.05$. The former might be compatible with a massive star progenitor if GRBs occur in the "superbubble" 
blown by preceding supernovae exploding in the same molecular cloud (Scalo \& Wheeler 2001), while the latter 
points to a progenitor with a low mass and low metallicity or, otherwise, to a small mass-loss rate and high
wind speed during the last few thousand years before the collapse, when the environment within 1 pc is shaped
by the stellar wind.

 As a final conclusion, we note that the Structured Outflow model is a serious contender to the Jet model 
in accommodating the broadband emission of GRB afterglows with optical light-curve breaks. In both models,
the best fit parameters describing the ejecta kinetic energy -- jet opening \& energy or core opening \& 
energy density along the outflow symmetry axis -- have narrow distributions, hinting to a possible universality 
of these parameters.

\begin{small}
\section*{Acknowledgments}
 This work was supported by NSF grant AST-0406878. \\
 Some or all of the radio data for the afterglows 011211, 020813, and 030226 referred to in this paper were 
 drawn from the GRB Large Program at the VLA \\ (http://www.vla.nrao.edu/astro/prop/largeprop) and can be found 
 at Dale Frail's website \\ http://www.aoc.nrao.edu/~dfrail/allgrb\_table.shtml .
\end{small}

\clearpage

\end{document}